\documentstyle[makeidx,pstricks,pst-node,epsf,amssymb,12pt]{article}

\input{tcilatex}

\begin{document}

\author{Massimo Di Pierro%
\footnote{Email address: {\tt mdp@fnal.gov}}  \\ \\ 
\it Fermi National Accelerator Laboratory \\
\it P.O. Box 500, Batavia, Illinois 60510}

\title{Matrix Distributed Processing:\\ A set of {\tt C++} Tools for 
implementing generic lattice computations on parallel systems}
\date{26 April 2001}
\maketitle

\begin{abstract}
We present a set of programming tools (classes and functions 
written in {\tt C++} and based on Message Passing Interface)
for fast development of generic parallel (and non-parallel)
lattice simulations. They are collectively called {\tt MDP~1.2}.

These programming tools include classes and algorithms for matrices,
random number generators, distributed lattices (with arbitrary topology), 
fields and parallel iterations. 
No previous knowledge of MPI is required in order to use them.

Some applications in electromagnetism, electronics, condensed matter
and lattice QCD are presented.
\end{abstract}

\newpage

{\LARGE Program Summary}

\begin{itemize}

\item {\it Title of the library:} {\tt MDP}, 
Matrix Distributed Processing (version 1.2)

\item {\it Computer for which the library is designed:} Any computer or parallel computer, 
including clusters of workstations.

\item {\it Software required in order to run MDP:} {\tt MPI} (Message Passing Interface) is required to run {\tt MDP} in parallel. {\tt MPI} is not required to run {\tt MDP} in single process mode.

\item {\it Computer(s) on which the library has been tested:}: SUN SparcSTATION with 
Solaris, Linux PCs, cluster of PCs (connected by Ethernet and Myrinet). 
The code has also been tested partially on a Cray T3E.

\item {\it Programming language used:} {\tt ANSI C++}, to be compiled with 
\begin{verbatim}
g++ -ansi -pedantic [input program] -o [output program]
\end{verbatim}

\item {\it No. of lines in distributed software, including test data, etc.:} 14372

\item {\it Nature of the physical problem:} Any problem that can be 
described in terms of interacting fields discretized on a 
lattice of some arbitrary shape and topology.

\item {\it Parallel applications provided together with the library as examples:} A program that solves electrostatic problems ({\tt application1.C}). A program that computes total impedance in a net of resistors ({\tt application2.C}). An Ising model simulation ({\tt application3.C}). A parallel implementations of the Vegas multidimensional integration algorithm ({\tt MDP\_PVegas.h}, not described here). A complete Lattice QCD package ({\tt FermiQCD}, not described here).

\item {\it Web site:} 

{\tt home.dencity.com/massimo\_dipierro/mdp.html}

\end{itemize}

\newpage
\tableofcontents
\newpage

\section{Introduction}

One of the biggest discoveries of modern physics is that long distance
forces are mediated by fields and these fields are subject to local
interactions. Local differential equations, in some field variables,
seem to describe any aspect of fundamental physics (from
particle physics to gravitation, from thermodynamics to condensed
matter).

Often the equations describing complex interacting systems cannot be
solved exactly and require a numerical approach, i.e. discretize  
on a lattice the space on which the equations are defined.
The derivative terms of a local equation, once discretized, give origin 
to quasi-local interaction terms on the lattice.

The good thing of having to deal with quasi-local equations is that one can
partition the problem by partitioning the lattice (the space) 
on which the problem is defined, and only a minimal set 
of communications is required to
pass information about the field variables at the boundary of these
partitions (because first and second derivatives of the 
differential equations in the continuum correspond to nearest neighbor
interaction terms on the lattice).

Parallel computers and clusters are the natural hardware for
implementing the numerical algorithms that can solve this kind of
problems.

{\tt MDP} is a set of {\tt C++} tools for fast development 
of any kind of parallel 
numerical computations that involve the lattice discretization 
of a finite portion of space (and/or time) over a number of parallel
processors (supporting Message Passing Interface). 
It considerably reduce that development costs since one can
concentrate on the algorithm instead of focusing on communications.
	
A number of tricks and optimizations are implemented
that make of {\tt MDP 1.2} an efficient 
package for developing, among others, fast and robust 
Lattice QCD applications\footnote{%
For example {\tt MDP 1.2} has been used here at Fermilab to develop,
in a relatively short time, a parallel software for Lattice QCD
simulations (see section~\ref{fermi}).}.

\subsection{Content}

The main characteristics of {\tt MDP 1.2} are:

\begin{itemize}
\item  It includes {\tt Matrix}, a class to manipulate 
matrices of complex (%
{\tt Complex}) numbers. 
The assignment operator and the copy constructor
have been overloaded to optimize the speed and the memory usage. In fact a
{\tt Matrix} object contains a dynamically allocated pointer to the
location of memory where the matrix elements are stored. The matrix elements
are automatically copied only when it is inevitable, otherwise their
memory address is copied. In particular, the temporary {\tt Matrix}
object returned by a function stores its elements in the same
location of memory used by the function that will take that object as
argument. This class is used to implement an interface to the fields.

\item  All the calls to MPI are done through a wrapper object, {\tt mpi},
which takes care of generating communication tags and computing the
length of the objects to be sent/received by the parallel
processes. 

\item The most important classes in {\tt MDP 1.2} are
\begin{verbatim} 
generic_lattice
site
generic_field
\end{verbatim}
They enable the user
to define lattices of any size, shape and topology and
associate any kind of structure to the sites. 
Other properties of these classes are: automatic 
communications between the parallel processes (when necessary), 
parallel I/O for the field variables and functions to move on the lattice.

\item  For each user defined lattice, one random generator per site is
automatically created and is a member variable of {\tt generic\_lattice}.
This guarantees that any simulation will give the same results whether it
runs on a single PC or in parallel, independently of the way the lattice is
partitioned. The random generator itself can be seen as a field and it 
can be saved and restored and any point.

\item  Each lattice (and each field defined on that lattice) can be
partitioned arbitrarily among the different processes (or automatically if
the user does not bother to do it) and the constructor of {\tt %
generic\_lattice}, for each process, takes care of allocating the memory for
storing the neighbor sites that have to be passed by a neighbor process.
The code has been optimized in order to minimize the need for communications
and the low level optimization tricks are completely hidden to the high
level programmer.

\item  The same code can run on single process on one PC or multiprocess
on a cluster of PCs (or on a supercomputer supporting MPI), without any
modification. The results are independent from the number of processes%
\footnote{%
Apart for rounding errors that usually are compiler and hardware dependent.}.
\end{itemize}

As an example of what one can do with {\tt MDP 1.2} we present a
parallel program that creates a field of $5\times 5$ matrices on a $4
\times 4 \times 4$ lattice (distributed on 4 processors) and sets each
matrix (for each site) 
equal to the inverse of the exponential of a random $SU(5)$ matrix (just to
make things a little complicated!). Than it averages each matrix with its 6
nearest neighbors.

\begin{verbatim}
00: // Program: example01.C
01; #define PARALLEL
02: #define Nproc 4
03: #include "MDP_Lib2.h"
04: #include "MDP_MPI.h"
05:
06: int main(int argc, char **argv) {
07:   mpi.open_wormholes(argc,argv);
08:   int               ndim=3; 
09:   int               mybox[]={4,4,4}; 
10:   generic_lattice   mylattice(ndim, mybox); 
11:   Matrix_field      F(mylattice,5,5);
12:   site              x(mylattice);
13:   forallsites(x)
14:     F(x)=inv(exp(mylattice.random(x).SU(5)));
15:   F.update();
16:   forallsites(x) {
17:     F(x)=1.0/7.0*(F(x)+F(x+0)+F(x-0)+F(x+1)+F(x-1)+F(x+2)+F(x-2)); 
18:     printf("F(x) for x=(%i,%i)\n", x(0), x(1));
19:     print(F(x));
20:   };
21:   mpi.close_wormholes();
22:   return 0;
23: };
\end{verbatim} 

Note that each line has a meaning and none of them is superfluous:

\begin{itemize}
\item  Line 1 tells that the job will run in parallel

\item  Line 2 tells that it will run on 4 processes (this statement is
optional since the number of processes can be determined at runtime)

\item  Line 3 and 4 includes the libraries of {\tt MDP 1.2}

\item  Line 7 establishes the communications among the processes

\item  Line 8 sets a variable containing the dimensions of the lattice

\item  Line 9 fills an array containing the size of the lattice

\item  Line 10 defines the lattice, {\tt mylattice}

\item  Line 11 defines a field of $5 \times 5$ matrices {\tt F} 
       on {\tt mylattice} ({\tt Matrix\_field} is built-in field based
	on {\tt generic\_field}).

\item  Line 12 defines a {\tt site} variable {\tt x} on {\tt mylattice}

\item  Lines 13,14 initialize each field variable {\tt F(x)} with the
inverse of the exponential of a random SU(5) (generated using the local
random generator of site {\tt x}).

\item  Line 15 performs all the communication to update boundary sites, i.e. sites that are shared by different processes. They are determined automatically and optimally when the lattice is declared.

\item  Lines 16,17 average F(x) with its 6 next neighbor sites

\item  Lines 18,19 print out the results

\item  Line 21 closes all the communications
\end{itemize}

The source code of {\tt MDP 1.2} is contained in the files
\begin{itemize}
\item {\tt MDP\_Lib2.h}
\item {\tt MDP\_MPI.h}
\item {\tt MDP\_ANSI.h}
\item {\tt MDP\_Measure.h}
\item {\tt MDP\_Fit.h}
\item {\tt MDP\_PVegas.h}
\item {\tt MDP\_Prompt.h}
\item {\tt MDP\_utils.h}
\end{itemize}
plus some examples and applications. Only the classes defined in the 
first two files will be discussed here.

A reader interested only in parallelization issues can jump directly to
section 4, and go back to sections 2 and 3 later. 

\section{Non Parallel Tools}

Four basic classes are declared in the file {\tt MDP\_Lib2.h} \footnote{%
The classes defined in the file {\tt MDP\_Lib2.C} have extensively
been used in some analysis programs written by the Southampton Theory
Group for Lattice QCD applications.}:

\begin{itemize}

\index{\tt myreal}
\index{\tt Complex}
\index{\tt DynamicArray<>}
\index{\tt Matrix}
\index{\tt Random\_generator}
\index{\tt JackBoot}

\item {\tt myreal}. It is equivalent to {\tt float} unless the user
defines the global constant {\tt USE\_DOUBLE\_PRECISION}. 
In this case {\tt myreal} stands for {\tt double}. 

\item  {\tt Complex}. Declared as {\tt complex\TEXTsymbol{<}myreal\TEXTsymbol{%
>}}. The imaginary unit is implemented as a global constant {\tt %
I=Complex(0,1)}.

\item {\tt DynamicArray<object,n>}. It is a container for {\tt n}-dimensional
arrays of {\tt object}s (for examples arrays of {\tt Matrix}). {\tt
DynamicArray}s can be resized any time, moreover they can be passed to
(and returned by) functions. 

\item  {\tt Matrix}. An object belonging to this class is effectively a
matrix of {\tt Complex} numbers and it may have arbitrary dimensions.

\item  {\tt Random\_generator}. This class contains the random number 
generator and
member functions to generate random {\tt float} numbers with uniform
distribution, Gaussian distribution or any user defined distribution. It can
also generate random $SU(n)$ matrices. Different {\tt Random\_generator}
objects can be declared and initialized with different seeds.

\item  {\tt JackBoot}. It is a class to store sets of data to be used
for computing the average of any user defined function acting on the
data. It includes member functions for computing
Jackknife and Bootstrap errors on this average.

\end{itemize}

Moreover the following constants are declared
\begin{verbatim}
#define and            &&
#define or             ||
#define Pi             3.14159265359
#define I              Complex(0,1)
#define PRECISION      1e-16
#define CHECK_ALL
#define TRUE  1
#define FALSE 0
\end{verbatim} 
If the definition of {\tt CHECK\_ALL} is removed the code 
will be faster but some checks will be skipped resulting a lesser safe code.

\subsection{On {\tt DynamicArray} and {\tt Matrix}}

The main idea on which these two classes have been built is the solution of the
problem of returning objects containing pointers to dynamically allocated
memory~\cite{core}.

Consider the following code involving matrices:
\begin{verbatim}
Matrix A,B(10,10);
A=B;
A=B*A;
\end{verbatim} 

In the first assignment one wants each element of {\tt B} to be copied in
the corresponding element of {\tt A}, while the second assignment is faster
if {\tt B} and {\tt A} are passed to the local variables of the {\tt %
operator*} by reference (i.e. without copying all the elements). Moreover
one wants the local variable created by the {\tt operator*} to occupy the
same memory location as the variable that is returned (this avoids copying
and wasting space). To implement this idea each {\tt Matrix} object contains
a {\tt FLAG} and a pointer to dynamically allocated memory (where the
numbers are stored). The copy constructor and the {\tt operator=} have been
overloaded in such a way to take care of the status of the {\tt FLAG} and
eventually to copy the pointer to the memory, instead of copying the memory
containing the real matrix.

A physical location of memory may be pointed by different {\tt Matrix}
objects, but this never generates confusion if a few safety rules,
stated later, are followed. An automatic system of garbage collecting
deallocates the unused memory when there are no objects alive pointing to it.

In this way, in the first assignment of the example, $11+800$ bytes\footnote{%
The number $11$ is the size in bytes of a {\tt Matrix} object. In is
independent from the size of the memory occupied by the ``real'' matrix.}
are copied, while in the second assignment only $11$ bytes are copied three
times (when passing {\tt A} and {\tt B} to the {\tt operator*()} and when
returning the result) to be compared with the $800$ bytes of the matrix. The
pointer of {\tt A} is swapped only when the multiplication is terminated
without generating confusion between the input and the output of the
function. The {\tt FLAG} takes care of everything and the procedure 
works also in any recursive expression.
For big matrices this is, in principle, 
faster than it would be possible in
FORTRAN or C. (In FORTRAN, for example, it would be necessary 
to create a temporary array where to store the result of the 
multiplication and copy it into {\tt A}).
In practice this is not true because the class {\tt Matrix} 
uses dynamic allocation and modern compilers are not able to optimize 
its algorithms as well as they can optimize algorithms using 
normal arrays.

We do not suggest to use the classes {\tt DynamicArray} and {\tt Matrix}
to implement those parts of the user's algorithms that are critical 
for speed.
Despite this, they can be used to 
implement an efficient and practical interface to field variables 
(as shown in subsection 4.3.4).

\index{safety rules}

\subsection{Safety rules}
The safety rules for {\tt DynamicArray} and {\tt Matrix} objects are:
{\bf \begin{itemize} 
\item[$\bigstar$]
Always, before returning an {\it object}, 
call {\tt prepare(}{\it object}{\tt);} 
 
\item[$\bigstar$]
Never explicitly call the copy constructor. 

\item[$\bigstar$]
Do not call the assignment operator of the argument of a function 
within the function itself.
\end{itemize}} 

\subsection{Using {\tt DynamicArray}}

\index{\tt DynamicArray<>}

DynamicArrays are declared using templates. For example the command
\begin{verbatim}
DynamicArray<float,4> myarray(2,3,6,5);
\end{verbatim} 
declares a $2 \times 3 \times 6 \times 5$ array of {\tt float} called
{\tt myarray}. The first argument of the template specifies the type
of object (i.e. {\tt float}) and the second argument is the number of
dimensions of the array (i.e. {\tt 4}).
An array of this kind can be resized at any time. For example
\begin{verbatim}
myarray.dimension(4,3,2,2);
\end{verbatim} 
transforms {\t myarray} in a new $4 \times 3 \times 2 \times 2$ array of {\tt
float}. A {\tt DynamicArray} can have up to 10 dimensions.
The array elements can be accessed by reference using {\tt operator()} in the
natural way
\begin{verbatim}
myarray(i,j,k,l)
\end{verbatim} 
where {\tt i,j,k,l} are integers.
The total number of dimensions (in this example 4) is contained 
in the member variable
\begin{verbatim}
myarray.ndim
\end{verbatim} 
Two useful member functions are
\begin{verbatim}
myarray.size()
\end{verbatim} 
that returns the total number of elements of the array (as {\tt long})
and
\begin{verbatim}
myarray.size(n)
\end{verbatim} 
that returns the size of the of the dimension {\tt n} of the array (as
{\tt long}). The argument {\tt n} must be an integer in the range
[0,{\tt ndim}-1]. 

DynamicArrays can be passed to function and can be returned by
functions but, before a DynamicArray is returned the function, 
{\tt prepare} must be called.
Here is an example of a program that returns a DynamicArray.
\begin{verbatim}
// Program: example02.C
#include "MDP_Lib2.h"

DynamicArray<float,2> f(float x) {
  DynamicArray<float,2> a(2,2);
  a(0,0)=x;
  prepare(a);  // IMPORTANT !!
  return a;
};

int main() {
  float x=f(Pi)(0,0);
  printf("%f\n", x);	
  return 0;
};
\end{verbatim} 
It prints {\tt 3.141593}. Note that {\tt f()} returns a $2 \times 2$
{\tt Matrix}.
The following is a program that contains a function that takes a
DynamicArray as argument 
\begin{verbatim}
// Program: example03.C
#include "MDP_Lib2.h"

DynamicArray<float,2> f(float x) {
  DynamicArray<float,2> a(2,2);
  a(0,0)=x;
  prepare(a);
  return a;
};

void g(DynamicArray<float,2> a) {
  printf("%f\n", a(0,0));
};

int main() {
  DynamicArray<float,2> a;
  a=f(Pi);
  g(a);
  return 0;
};
\end{verbatim} 
The output of this program is again {\tt 3.1415}. Note that when {\tt
f()} is assigned to {\tt a}, the latter is automatically resized.

Bare in mind that even if the function {\tt g()} does not take its
arguments by reference the object {\tt DynamicArray} contains a pointer
to the memory where the real data are stored, therefore it is {\it as if} it
was passed by reference. This does not generate confusion providing
one follows the three safety rules.

\subsection{Using {\tt Matrix}}

\index{\tt Matrix}

A {\tt Matrix} object, say {\tt M}, is very much like {\tt
DynamicArray<Complex,2>}. 
A {\tt Matrix} can be declared either by specifying its size
or not
\begin{verbatim}
Matrix M(r,c);     // r rows times c columns
Matrix M;          // a general matrix
\end{verbatim} 

\noindent Even if the size of a matrix has been declared it can be changed
anywhere with the command
\begin{verbatim}
M.dimension(r,c);  // r rows times c columns
\end{verbatim} 

\noindent Any {\tt Matrix} is automatically resized, if necessary, when a
value is assigned to it. The following lines 
\begin{verbatim}
Matrix M(5,7), A(8,8);
M=A;
printf("%i,%i\n", M.rowmax(),M.colmax());
\end{verbatim} 
prints {\tt 8,8},
\noindent The member functions {\tt rowmax()} and {\tt colmax()} return
respectively the number of rows and columns of a {\tt Matrix}.

The element $(i,j)$ of a matrix {\tt M} can be accessed with the natural
syntax
\begin{verbatim}
M(i,j)
\end{verbatim} 
where {\tt i,j} are integers.
\noindent Moreover the class contains functions to perform standard
operations among matrices:
\begin{verbatim}
+, -, *, /, +=, -=, *=, /=, inv, det, exp, sin, cos, log,
transpose, hermitiam, minor, identity,...
\end{verbatim} 

\noindent As an example a program to compute

\begin{equation}
\left[ \exp \left( \matrix{ 2 & 3 \cr 4 & 5\mathit{i} } \right) \right]^{-1}
\end{equation}
looks like this:
\begin{verbatim}
// Program: example04.C
#include "MDP_Lib2.h"  
int main() {
   Matrix a(2,2);
   a(0,0)=2; a(0,1)=3;
   a(1,0)=4; a(1,1)=5*I;
   print(inv(exp(a)));
   return 0;
};
\end{verbatim} 

\noindent It is straightforward to add new functions copying the
following prototype 
\begin{verbatim}
Matrix f(const Matrix a, const Matrix b, ...) {  
   Matrix M;
   // body,
   prepare(M);
   return M;
};  
\end{verbatim} 

We repeat again the three safety rules one should always remember: 
one should call {\tt prepare()}
before returning a {\tt Matrix}; one should not return a {\tt Matrix} by
reference; one should not initialize a {\tt Matrix} using the copy
constructor nor call the assignment operator of a {\tt Matrix} argument.

Here is one more example of how to use the class {\tt Matrix}:

\begin{verbatim}
// Program: example05.C
#include "MDP_Lib2.h"  
Matrix cube(Matrix X) {  
   Matrix Y;
   Y=X*X*X;
   prepare(Y);
   return Y;
};
int main() {
   Matrix A,B;
   A=Random.SU(3);
   B=cube(A)*exp(A)+inv(A);
   print(A);
   print(B);
   return 0;
};
\end{verbatim} 

\noindent This code prints on the screen a random SU(3) matrix $A$ and $%
B=A^3e^A+A$.\newline
Some example statements are listed in tab.~\ref{figmatrix1}. Note that
the command
\begin{verbatim}
A=mul_left(B,C);
\end{verbatim} 
is equivalent but faster than
\begin{verbatim}
A=C*transpose(B);
\end{verbatim} 
because it does not involve allocating memory for the transposed of
{\tt B}.

\begin{table}
\begin{center}
\begin{tabular}{|l|l|} 
\hline 
Example & {\tt C++ with MDP\_Lib2.h} \\ \hline 
$A\in M_{r\times c}({\bf C})$ & {\tt A.dimension(r,c)} \\  
$A_{ij}$ & {\tt A(i,j)} \\ 
$A=B+C-D$ & {\tt A=B+C-D} \\
$A^{(ij)}=B^{(ik)}C^{(kj)}$ & {\tt A=B*C} \\ 
$A^{(ij)}=B^{(jk)}C^{(ik)}$ & {\tt A=mul\_left(B,C)} \\ 
$A=aB+C$ & {\tt A=a*B+C} \\ 
$A=a{\bf 1}+B-b{\bf 1}$ & {\tt A=a+B-b} \\ 
$A=B^TC^{-1}$ & {\tt A=transpose(B)*inv(C)} \\ 
$A=B^{\dagger }\exp (iC)$ & {\tt A=hermitian(B)*exp(I*C)} \\
$A=\cos (B)+i\sin (B)*C$ & {\tt A=cos(B)+I*sin(B)*C} \\
$a=\func{real}(tr(B^{-1}C))$ & {\tt a=real(trace(inv(B)*C))} \\
$a={\det (B) det (B}^{-1})$ & {\tt a=det(B)*det(inv(B))} \\ \hline 
\end{tabular} 
\caption{Examples of typical instructions acting on {\tt Matrix}
objects.  {\tt A,B,C,D} are assumed to be declared 
as {\tt Matrix}; {\tt r,c} as {\tt int}; {\tt a,b} may be any kind of
number.} 
\label{figmatrix1}
\end{center}
\end{table}

\index{\tt DynamicArray<Matrix,>}

As one more example of a powerful application of {\tt DynamicArray} 
here is a program that defines a multidimensional {\tt DynamicArray}
of {\tt Matrix} objects and pass it to a function. 

\begin{verbatim}
// Program: example06.C
#include "MDP_Lib2.h"

DynamicArray<Matrix,3> initialize() {
  DynamicArray<Matrix,3> d(20,20,20);
  int i,j,k;	
  for(i=0; i<20; i++)
    for(j=0; j<20; j++)
      for(k=0; k<20; k++) {
        d(i,j,k).dimension(2,2);	
        d(i,j,k)(0,0)=k;
        d(i,j,k)(0,1)=i;
        d(i,j,k)(1,0)=j;
        d(i,j,k)(1,1)=k;
      };
  prepare(d);
  return(d);
}

DynamicArray<Matrix,3> f(DynamicArray<Matrix,3> c) {
  DynamicArray<Matrix,3> d(c.size(0),c.size(1),c.size(2));
  int i,j,k;
  for(i=0; i<c.size(0); i++)
    for(j=0; j<c.size(1); j++)
      for(k=0; k<c.size(2); k++)
        d(i,j,k)=sin(c(i,j,k));
  prepare(d);
  return(d);
};

int main() {
  DynamicArray<Matrix,3> a, b;
  a=initialize();
  b=f(a);

  int i=1, j=2, k=3;
  print(a(i,j,k));
  print(b(i,j,k));
  return 0;
};
\end{verbatim} 

{\tt a}, {\tt b}, {\tt c}, {\tt d} are 3D $20 \times 20 \times
20$ arrays of $2 \times 2$ matrices. The program prints
\begin{verbatim}
[[ 3.000+0.000i  1.000+0.000i ] 
 [ 2.000+0.000i  3.000+0.000i ]]
[[ 0.022+0.000i  -0.691+0.000i ] 
 [ -1.383+0.000i  0.022+0.000i ]]
\end{verbatim} 
Note that the instruction
\begin{verbatim}
d(i,j,k)=sin(c(i,j,k));
\end{verbatim} 
computes the {\tt sin()} of the {\tt Matrix} {\tt c(i,j,k)}. 

\subsection{\tt class Random\_generator}

\index{\tt Random\_generator}
\index{random {\tt plain}}
\index{random {\tt gaussian}}
\index{random {distribution}}
\index{random $SU(n)$}

The random generator implemented in {\tt Random\_generator} is the Marsaglia
random number generator described in ref.\cite{marsaglia} 
(the same generator is also used by
the UKQCD collaboration in many large scale numerical simulations). It
presents the nice feature that it can be initialized with a single long
random number and its correlation time is relatively large
compared with the standard random generator {\tt rand()} of C++. To define a
random generator, say {\tt myrand}, and use the integer {\tt seed} as 
seed one should simply pass {\tt seed} to the constructor:

{\tt Random\_generator myrand(seed);}

For simple non-parallel applications one only needs one random generator.
For this reason one {\tt Random\_generator} object called {\tt Random} is
automatically created and initialized with seed 0.

{\tt Random\_generator} contains some member variables for the seeds and
four member functions:

\begin{itemize}
\item  {\tt plain();} It returns a random {\tt float} number in the
interval [{\tt 0},1) with uniform distribution.

\item  {\tt gaussian();} It returns a random {\tt float} number $x$
generated with a Gaussian probability $P(x)=\exp (-x^2/2).$

\item  {\tt distribution(float (*P)(float, void*), void* a);} It
returns a random number $x$ in the interval [0,1) generated with a Gaussian
probability {\tt P(x,a)}, where {\tt a} is any set of parameters pointed by 
{\tt void} *{\tt a} (the second argument passed to {\tt distribution}). The
distribution function $P$ should be normalized so that $0\leq \min P(x)\leq 1
$ and $\max P(x)=1.$

\item  {\tt Random\_generator::SU(int n);} it returns a random {\tt %
Matrix} in the group $SU(n)$ or $U(1)$ if the argument is $n=1$.
\end{itemize}

Here is a simple program that creates one random number generator, generates a
set of 1000 random Gaussian numbers and counts how many of them are in the
range $[n/2,(n+1)/2)$ for $n=0...9$
\begin{verbatim}
// Program: example07.C
#include "MDP_Lib2.h"
int main() {
   Random_generator random;
   int i,n,bin[10];
   float x;
   for(n=0; n<10; n++) bin[n]=0;
   for(i=0; i<1000; i++) {
      x=random.gaussian();
      for(n=0; n<10; n++)
         if((x>=0.5*n) && (x<0.5*(n+1))) bin[n]++;
   };
   for(n=0; n<10; n++) 
     printf("bin[%i] = %i\n", n, bin[n]);
   return 0;
};

\end{verbatim} 

Here is a program that computes the average of the sum of two sets of
1000 numbers, {\tt a} and {\tt b}, where {\tt a} is generated with 
probability $P(a)=\exp (-(a-\bar a)^2/(2\sigma))$ and {\tt b} with
probability $Q(b)=\sin(\pi b)$.
\begin{verbatim}
// Program: example08.C
#include "MDP_Lib2.h"

float Q(float x, void *a) {
   return sin(Pi*x);
};

int main() {
   Random_generator random;
   int i,n,bin[10],N=100;
   float a,b,average=0, sigma=0.3, a_bar=1;
   for(i=0; i<N; i++) {
      a=(sigma*random.gaussian()+a_bar);
      b=random.distribution(Q);
      average+=a+b;
      printf("average=%f\n", average/(i+1));
   };
   return 0;
};
\end{verbatim} 
For large {\tt N} the output asymptotically approaches {\tt a\_bar+0.5}=1.5.

The algorithm for SU(2) is based on map between $O(3)$ and $SU(2)$, realized
by 
\begin{equation}
\{\widehat{a},\alpha \}\rightarrow \exp (i\alpha \widehat{a}\cdot \sigma
)=\cos (\alpha )+i\widehat{a}\cdot \sigma \sin (\alpha )
\end{equation}
where ($\sigma ^1,\sigma ^2,\sigma ^3$) is a vector of Pauli matrices,  
$\widehat{a}\in S^2$ is a uniform vector on the sphere and 
$\alpha \in [0,\pi )$ is a uniform rotation angle
around that direction.

A random $SU(n)$ is generated using the well-known  
Cabibbo-Marinari iteration ~\cite{cm} of the algorithm for SU(2).

\subsection{{\tt class JackBoot }}

\index{\tt JackBoot}
\index{JackKife error}
\index{Bootstrap error}

Suppose one has $n$ sets of $m$ different measured {\tt float} 
quantities {\tt x[i][j]}
(where $i=0...n$ and $j=0...m$) and one wants to compute the average over $i$%
, of a function {\tt F(x[i])}. For example
\begin{verbatim}
float x[n][m]={....}
float F(float *x) {
   return x[1]/x[0];
};
float result=0;
for(i=0; i<n; i++)
   result+=F(x[i])/n;
};
\end{verbatim} 

Than one may ask: what is the error on the result? 
In general there are two algorithms to estimate this error:
Jackknife and Bootstrap~\cite{errors}. They are both implemented as member
functions of the class JackBoot. They work for any arbitrary function F.

A {\tt JackBoot} object, let's call it {\tt jb}, is a sort of container for
the data. After it has been filled it can be asked to return the mean of 
{\tt F()} and its errors. Here is an example of how it works:
\begin{verbatim}
JackBoot jb(n,m); 
jb.f=F;
for(i=0; i<n; i++)
  for(j=0; j<m; j++)
     jb(i,j)=x[i][j]; 
printf("Result          = %f\n", jb.mean()); 
printf("Jackknife error = %f\n", jb.j_err()); 
printf("Bootstrap error = %f\n", jb.b_err(100));
\end{verbatim} 
Note that
\begin{itemize}
\item  The constructor of the class {\tt JackBoot} takes two
arguments: the first is the number of configuration; the second is
the number of the quantities measured on each configuration.

\item  {\tt jb.f} is the pointer to the function used in the analysis.

\item  {\tt jb.mean()} returns the mean.

\item  {\tt jb.j\_err()} returns the Jackknife error.

\item  {\tt jb.b\_err()} returns the Bootstrap error. It takes as argument
the number of Bootstrap samples. The default value is 200.
\end{itemize}

It is possible to declare arrays of {\tt JackBoot} objects, but it is rarely
necessary. It is simpler to declare different functions and repeat the
analysis using the same {\tt JackBoot} object assigning the pointer {\tt %
JackBoot::f} to each of the functions at the time. \newline
As another example consider the following program. It generates an array of
100 $SU(6)$ matrices. For each matrix it computes trace and determinant, and
returns the average of the ratio between the real part of the trace and the
real part of the determinant (with its Jackknife and Bootstrap errors), and the average of the product of the real part of the trace and the
real part of the determinant (with its Jackknife and Bootstrap errors)
\begin{verbatim}
// Program: example09.C
#include "MDP_Lib2.h" 
#define n 100 
float f1(float *x, float *a) { return x[0]/x[1]; };
float f2(float *x, float *a) { return x[0]*x[1]; };
int main() {
   Matrix A;
   JackBoot jb(n,2);
   int i;
   for(i=0; i<n; i++) { 
      A=Random.SU(6);
      jb(i,0)=real(det(inv(A))); 
      jb(i,1)=real(det(A)); 
   };
   jb.f=f1;
   printf("Result x[0]/x[1] = %f\n", jb.mean());
   printf("Jackknife error  = %f\n", jb.j_err());
   printf("Bootstrap error  = %f\n", jb.b_err(100));
   jb.f=f2;
   printf("Result x[0]*x[1] = %f\n", jb.mean());
   printf("Jackknife error  = %f\n", jb.j_err());
   printf("Bootstrap error  = %f\n", jb.b_err(100));
   return 0;
};
\end{verbatim} 

Note that any user defined function used by {\tt JackBoot} 
({\tt f1} and {\tt f2} in the example) 
has to take two arguments: an array of {\tt float}
(containing a set of measurements done on a single configuration) and a
pointer to {\tt void}. This is useful to pass some extra data to the
function.
The extra data must be passed to {\tt JackBoot} by assigning the
member variable
\begin{verbatim}
void* JackBoot::handle;
\end{verbatim} 
\noindent to it.

JackBoot has one more member function {\tt plain(int i)}, where the call
\begin{verbatim}
/* define JackBoot jb(...) */
/* assign a value to the integer i */
jb.plain(i)
\end{verbatim} 
is completely equivalent to
\begin{verbatim}
/* define JackBoot jb(...)          */
/* assign a value to the integer i */
float f(float *x, void *a) {
  return x[i];
};
jb.f=f
\end{verbatim} 
Using {\tt plain} saves one from defining trivial functions for JackBoot.

\section{Parallel Tools}

\index{Message Passing Interface (MPI)}

With a parallel program we mean a job constituted by many programs running
in parallel (eventually on different processors) to achieve a global goal.
Each of the running programs, part of the same job, is called a process.
Different processes can communicate to each other by exchanging messages (in
MPI) or by accessing each-other memory (in computers with shared
memory). The latter case is hardware dependent and will not be taken in
consideration. With the term ``partitioning'' we will refer to the way
different variables (in particular the field variables defined on a
lattice) are distributed among the different processes.

The most common and portable parallel protocol is Message Passing Interface
(MPI). It is implemented on a number of different platforms, for example
Unix, Linux, Solaris, Windows NT and Cray T3D/E. 

When writing a program using MPI, one 
essentially writes one program for each processor 
(Multiple Instructions Multiple Data) and MPI provides the
functions for sending/receiving information among them.

We present here an example of how MPI works, even if 
a detailed knowledge of MPI is not required to understand 
the rest of this paper.

Suppose one wants to compute $(5*7)+(4*8)$ in parallel using two processes.
A typical MPI program to do it is the following:

\begin{verbatim}
// Program: example10.C
#include "stdio.h" 
#include "mpi.h"
int main(int argc, char **argv) {
   int ME, Nproc;
   MPI_Status status;
   MPI_Init(&argc, &argv);
   MPI_Comm_rank(MPI_COMM_WORLD, &ME);
   MPI_Comm_size(MPI_COMM_WORLD, &Nproc);
   if(ME==1) {
      int b;
      b=4*8;
      MPI_Send(&b,1,MPI_INT,0,45,MPI_COMM_WORLD);
   };
   if(ME==0) {
      int a,b;
      a=5*7;
      MPI_Recv(&b,1,MPI_INT,1,45,MPI_COMM_WORLD,&status);
      printf("%i\n", a+b);
   };
   MPI_Finalize();
};
\end{verbatim} 

{\tt ME} is a variable that contains the unique identification number 
of each running process and {\tt Nproc} is the total number of
processes running the same code. 
Since different processes have different values of {\tt ME} they
execute different parts of the program ({\tt if(ME==n) {...}}).

In the example $(5*7)$ is computed by process 0 while, at the same time $(4*8)$
is computed by process 1. Then process 1 sends ({\tt MPI\_Send}) its partial
result to process 0, which receives it ({\tt MPI\_Recv}) and prints out the
sum\footnote{%
Needless to say that in such a simple program the parallelization is
completely useless because the message passing takes more time than the
computation of the whole expression!}. 

MPI instructions can be classified in

\begin{itemize}
\item  initialization functions (like {\tt MPI\_Init}, {\tt MPI\_Comm\_rank}%
, {\tt MPI\_Comm\_size})

\item  one to one communications (like {\tt MPI\_Send}, {\tt MPI\_Recv})

\item  collective communication (where for example one variable can be
broadcasted by one process to all the others)
\end{itemize}

It is not a purpose of this paper to explain how MPI works therefore we
refer to~\cite{mpi}.

\subsection{{\tt mpi} on top of {\tt MPI}}

\index{\tt mpi}
\index{identification number}
\index{\tt ME}
\index{\tt Nproc}
\index{\tt PARALLEL}

MPI is not Object Oriented and it is so general that
MPI function calls usually require many arguments.
This makes it very easy to do mistakes when programming with MPI. 

\begin{figure}
\epsfxsize=10cm
\epsfysize=10cm
\hfil \epsfbox{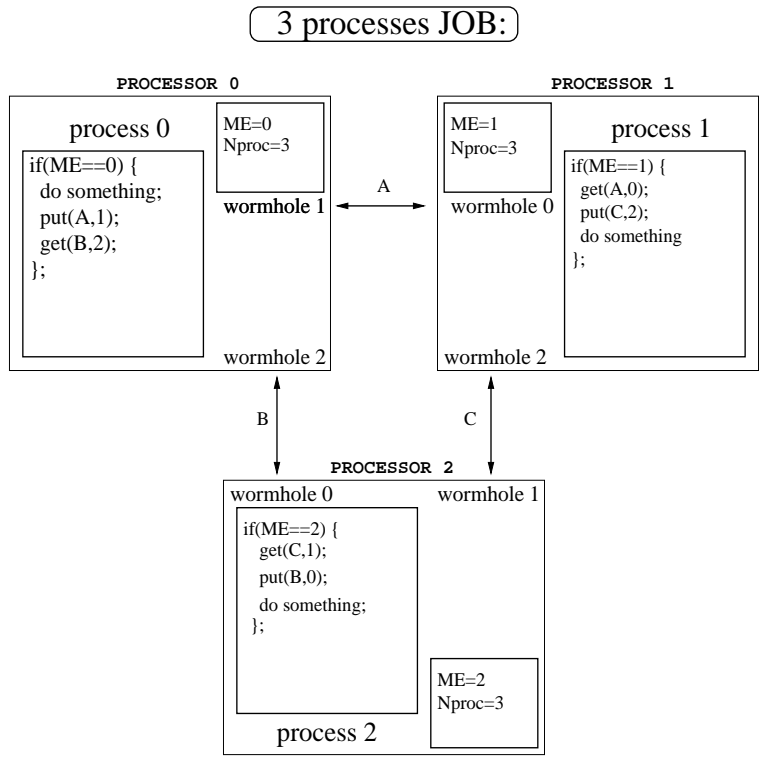} \hfil
\caption{Example of a 3 processes jobs. ME and Nproc are the global
variables that characterize the different processes running
(eventually) on different processors. The term wormhole refers to the
virtual connection between each couple of processes.}
\label{scheme}
\end{figure}

For this reason the class {\tt mpi\_wormhole\_class}, a wrapper to MPI, has
been created. It has only one global representative object called {\tt mpi}.
We believe that {\tt mpi} is easier to use than MPI in the context of
lattice simulation. 

{\tt mpi} is not intended as a substitute or an extension to {\tt
MPI}, in fact only a subset of the features of MPI
are implemented in {\tt mpi} and occasionally one may need explicit
{\tt MPI} calls. Its main purpose is to build an intermediate
programming level between {\tt MPI} and {\tt generic\_lattice}/{\tt
generic\_field}, which constitute the very essence of {\tt MDP 1.2}.
There are some advantages of having this intermediate level, for example:
\begin{itemize}
\item It allows one to compile the code without linking with {\tt MPI}.
In this case {\tt mpi} substitutes some dummy functions 
to the {\tt MPI} calls. This allows one to compile and run
any parallel program on a single processor computer even if {\tt MPI} is
not installed, and it is also important for testing and debugging the
code.

\item In case one wants to run {\tt MDP 1.2} on a parallel machine that
does not support {\tt MPI} but does support a different protocol, one
does not have to rewrite everything but only minor modifications to
the class {\tt mpi\_wormhole\_class} will be necessary.
\end{itemize}

\subsection{{\tt mpi} and Wormholes}

\index{wormholes}
\index{\tt mpi\_wormhole\_class}
\index{\tt mpi.open\_wormholes()}
\index{\tt mpi.close\_wormholes()}

From now one we will use here the word ``wormhole'', in its intuitive sense,
to represent the virtual connections
between the different processes. One can think of each process to have one
wormhole connecting it with each of the other processes. Each wormhole has a
number, the number of the process connected to the other end. Process X can
send something to process Y just putting something into its wormhole n.Y.
Process Y has to get that something from its wormhole n.X. {\tt mpi} is the
global object that allows the user to access these wormholes (put/get
stuff to/from them). A schematic representation of a 3 processes job is
shown in fig.~\ref{scheme}.

Using {\tt mpi}, instead of MPI, program {\tt example10.C} becomes:
\begin{verbatim}
// Program: example11.C
#define PARALLEL
#define Nproc 2
#include "MDP_Lib2.h"
#include "MDP_MPI.h" 
int main(int argc, char **argv) {
   mpi.open_wormholes(argc,argv);
   if(ME==1) {
      int b;
      b=4*8;
      mpi.put(b,0);
   };
   if(ME==0) {
      int a,b;
      a=5*7;
      mpi.get(b,1);
      printf("%i\n" , a+b);
   };
   mpi.close_wormholes();
};
\end{verbatim} 

The most important things to notice are:

\begin{enumerate}
\item  A constant (without any value assigned), {\tt PARALLEL}, is defined
on top. It tells the compiler to use the MPI functions. When it is omitted
some dummy functions are substituted to the MPI calls and communications are
skipped. This is useful for testing the behavior of a particular process in
a multiprocess job without actually going parallel.

\item  {\tt Nproc} is the total number of processes. 
If it is not defined by the user, {\tt Nproc} it is defined
automatically (as a variable) when {\tt MDP\_MPI.h} is included 
and its value is determined at runtime.

\item  Two libraries {\tt MDP\_Lib2.h} and {\tt MDP\_MPI.h} are included.
The standard headers (stdio.h, math.h, complex.h, iostream.h) are included
by {\tt MDP\_Lib2.h}.

\item  {\tt ME} is a global macro meaning {\tt mpi.my\_id}, i.e. the
identification number of the running process. Its value is assigned by the
call to {\tt mpi.open\_wormholes} when running with {\tt PARALLEL}.
Otherwise is has to be specified by the user after that call.

\item  All of the initialization functions have been replaced by a single
one: 
\begin{verbatim}
void mpi_wormhole_class::open_wormholes(int&,char**);
\end{verbatim} 
\noindent It is compulsory even if the program does not run in
parellel. It takes as input the same arguments of {\tt main()}%
\footnote{%
This is due to the fact that when running in parallel (using the {\tt runmpi}
command) a script submits the same program to the different processors with
different command parameters, which are passed as arguments to {\tt main()}.
This is how a process knows its own identification number: {\tt ME}.}.

\item  {\tt MPI\_Send} and {\tt MPI\_Recv} have been replaced by {\tt mpi.put%
} and {\tt mpi.get} respectively. They check automatically the size and type
of their arguments.

\item  At the end of the program one has to {\tt close\_wormholes}. Actually
if one forgets about this the destructor of {\tt mpi} does take care
of it.
\end{enumerate}

The programmer is forced to use the names {\tt Nproc} and {\tt ME} for the
total number of processes and the process identification number
respectively. They are keywords for {\tt MDP 1.2}.

\subsubsection{{\tt get()} and {\tt put()}}

\index{\tt mpi.put()}
\index{\tt mpi.get()}

The functions put/get have the following syntax
\begin{verbatim}
template <class T> mpi_wormhole_class::put(T&,int);
template <class T> mpi_wormhole_class::put(T&,int,mpi_request&);
template <class T> mpi_wormhole_class::get(T&,int)
\end{verbatim} 

The first argument (passed by reference) is the object to be sent ({\tt put})
or where to store the received one ({\tt get}). The second argument is the
destination ({\tt put}) or the source ({\tt get}). Forget about {\tt %
mpi\_request }for the moment, it is optional.

All the variables related with message passing are members of{\tt \ mpi}. In
this way variables that are not explicitly necessary remain hidden inside it
(for example the communicator and the communication tag).

The functions {\tt put } and {\tt get} that replaced {\tt MPI\_Send} and 
{\tt MPI\_Recv} are Object Oriented in the sense that they can be used to
send any kind of object\footnote{%
Assuming it does not contain a pointer do dynamically allocated memory.}.
The size and the tag of the communication are automatically computed and do
not have to be specified by the programmer. For example with the same syntax
a {\tt Complex} number can be put/get instead of an integer. The only price
paid for this is that a process, X, cannot send more that one object to
another process, Y, until the first object has been received. This is not
such a price because in the kind of computations we are interested in, we
want to minimize the number of put/get and we want to pack as much data as
possible in each single communication, instead of making many small ones.

To avoid confusion, if the third argument of {\tt put()} is not specified
this function will not return until the send process has been completed
(synchronous send)\footnote{%
Actually in {\tt mpi.put} an synchronous send is internally implemented using 
{\tt MPI\_Isend} + {\tt MPI\_Wait}. This is because {\tt MPI\_Send} is
ambiguous in some cases: the same program can have different behavior
depending on the size of the objects that are passed. Our implementation
avoids any ambiguity.}. Sometimes one wants the {\tt put()} command to
return even if the send is not completed, so that the process is free to do
more computations, and eventually check at a later stage if the previous 
{\tt put()} has completed its task (asynchronous send). One typical case is
when the object to be send was dynamically allocated and it cannot be
deallocated until the communication is concluded. The right way to implement
this send is writing a code like the following:
\begin{verbatim}
/* mystruct is any built in variable
   or user defined structure */
mystruct *myobj=new mystruct;
mpi_request request;
put(*myobj,destination,request);
/* do something else */
mpi.wait(request); 
delete myobj;
\end{verbatim} 

The {\tt wait()} member function of {\tt mpi} stops the process until the 
{\tt put()} command has terminated its task. Its only argument is the object
belonging to the class {\tt mpi\_request} that was passed to the
corresponding {\tt put()}. The function {\tt wait()} can also take a pointer
to an array of requests and wait for all of them. The general definition of
this function is
\begin{verbatim}
void mpi_wormhole_class::wait(mpi_request&);
void mpi_wormhole_class::wait(mpi_request*, int);
\end{verbatim} 

\noindent the second argument is the length of the array.

It is also possible to put/get arrays of objects by using the overloaded
functions
\begin{verbatim}
template <class T> mpi_wormhole_class::put(*T,long,int)
template <class T> mpi_wormhole_class::put(*T,long,int,mpi_request&)
template <class T> mpi_wormhole_class::get(*T,long,int)
\end{verbatim} 

\noindent where the first argument is a pointer to the first element of the
array, the second is the length, the third is the destination/source.

\subsubsection{put/get matrices}

\index{{\tt Matrix}! {\tt put()/get()}}

We have said that with the put/get commands it is not possible to pass
objects that contain a pointer to dynamically allocated memory and this is
the case for a Matrix object. This does not mean that one cannot put/get a
Matrix. The correct way to send a Matrix is
\begin{verbatim}
Matrix A;
/* do something with A */
mpi.put(A.address(),A.size(),destination);
\end{verbatim} 

The correct way to receive a Matrix is
\begin{verbatim}
Matrix B;
/* dimension B as A */
mpi.get(B.address(),B.size(),destination);
\end{verbatim} 

It is extremely important to stress that when one puts/gets a matrix one
only sends/receives its elements. Therefore (in the last example) {\tt B}
must have the same dimensions of {\tt A} at the moment {\tt get} is called,
otherwise one gets wrong results.

\subsubsection{Global communications }

\index{\tt mpi.add()}
\index{\tt mpi.barrier()}
\index{\tt mpi.abort()}
\index{\tt mpi.broadcast()}

As an example of global communication we will consider the following code
\begin{verbatim}
// Program: example12.C
#define PARALLEL
#define Nproc 2
#include "MDP_Lib2.h" 
#include "MDP_MPI.h" 
int main(int argc, char **argv) {
   mpi.open_wormholes(argc,argv);
   int a;
   if(ME==1) {
     a=4*8;
     mpi.add(a);
   };
   if(ME==0) {
     a=5*7;
     mpi.add(a);
     printf("%i\n",a);
   };
   mpi.close_wormholes();
};
\end{verbatim} 

The member function:
\begin{verbatim}
void mpi_wormhole_class::add(float&);
\end{verbatim} 

\noindent sums the first argument passed by all the processes. 
In this way each process
knows the result of the sum operation.

To sum an array of float the corresponding sum for arrays of float is
implemented as:
\begin{verbatim}
void mpi_wormhole_class::add(float*,long);
\end{verbatim} 

\noindent (the second argument is the length of the array). The same function 
{\tt add} also works for double, long, int, Complex and Matrix and arrays of 
these types.
To add {\tt Complex} numbers:
\begin{verbatim}
Complex x;
mpi.add(x);
\end{verbatim} 
To add {\tt Matrix} objects 
\begin{verbatim}
Matrix A;
mpi.add(A);
\end{verbatim} 

Some more member functions of {\tt mpi} for collective communications are:
\begin{itemize}

\item {\tt barrier()} which sets a barrier and all processes stop at that
point until all the processes reach the same point. 

\item {\tt abort()} which forces all the processes to abort.

\item {\tt template<class T> broadcast(T a, int p)} which broadcast the
object {\tt a} (belonging to an arbitrary class {\tt T}) of process
{\tt p} to all other processes. 

\item {\tt template<class T> broadcast(T *a, long n, int p)} which broadcast
the array {\tt a} of objects {\tt T}  of process {\tt p} to all other
processes. {\tt n} here is the number of elements of the array.

\end{itemize}

Here is an example of how to use these collective communications
\begin{verbatim}
// Program: example13.C
#define PARALLEL
#define Nproc 2
#include "MDP_Lib2.h" 
#include "MDP_MPI.h" 
int main(int argc, char **argv) {
   mpi.open_wormholes(argc,argv);
   int i,j;
   for(i=0; i<5; i++) {
      if(ME==0) j=i;
      else      j=0;
      if(i%2==0) mpi.barrier();
      mpi.broadcast(j,0);
      printf("I am process %i, i=%i, j=%i\n", ME, i,j);
   };
  mpi.close_wormholes();
};
\end{verbatim} 
This programs prints
\begin{verbatim}
I am process 0, i=0, j=0
I am process 0, i=1, j=1
I am process 0, i=2, j=2
I am process 0, i=3, j=3
I am process 0, i=4, j=4
I am process 1, i=0, j=0
I am process 1, i=1, j=1
I am process 1, i=2, j=2
I am process 1, i=3, j=3
I am process 1, i=4, j=4
\end{verbatim} 

\section{ Introducing Lattices and Fields}

\subsection{\tt class generic\_lattice}

\index{lattice!define}
\index{lattice!dimensions}
\index{lattice!box}
\index{Lattice!partitioning}
\index{Lattice!topology}
\index{\tt generic\_lattice}

By a lattice we mean a generic set of points and a set of functions to move
from one point to another. In {\tt MDP 1.2} a lattice is implemented as
any subset of a regular n-dimensional grid with a given (arbitrary) topology.

A lattice of such a kind is implemented as
an object belonging to the class {\tt generic\_lattice} using the command:
\begin{verbatim}
generic_lattice mylattice(ndim,mybox,mypartitioning,
                          mytopology,seed,next_next);
\end{verbatim} 
or the command
\begin{verbatim}
generic_lattice mylattice(ndim,ndir,mybox,mypartitioning,
                          mytopology,seed,next_next);
\end{verbatim} 
\noindent where:

\begin{itemize}
\item  {\tt mylattice} is a user defined variable that will contain the
actual lattice structure.

\item  {\tt ndim} is the dimension of the basic regular grid.

\item  {\tt ndir} is the number of directions. It can be omitted and, by
default, it is assumed to be equal to {\tt ndim}.

\item  {\tt mybox} is a user defined {\tt ndim}-ensional array of {\tt
int} that contains the size of the basic regular grid.

\item  {\tt mypartitioning} is a user defined function that for each site of
the basic grid returns the number of the process that stores it (or it
returns {\tt NOWHERE} if the point has to be excluded from the
lattice). If {\tt mypartitioning} is omitted all
sites of the original grid (specified by mybox) will be part of the final
lattice, and they will be equally distributed between the processes
according with the value of coordinate 0 of each point\footnote{
For example, consider a lattice of size 8 in the 0 direction
distributed on 4 processes. Process 0 will contain sites coordinate 0
equal to 0 and 1. Process 1 will contain sites coordinate 0
equal to 2 and 3. An so on.}.

\item  {\tt mytopolgy} is a user defined function that for each site of the
final (remaining) grid, and for each dimension in the {\tt mydim}-ensional
space, returns the coordinates of the neighbor sites in the up and down
direction. If {\tt mytopology} is omitted the lattice will, by
default, have the topology of an {\tt ndim}-ensional torus, i.e. the
same of the basic grid plus periodic boundary conditions).

\item {\tt seed} is an integer that is used to construct a site
dependent seed to initialize the random generators
associated to each lattice sites. If omitted it is assumed to be 0.

\item {\tt next\_next} is an integer that may have only three values
(1,2 or 3). It essentially fixes the thickness of the boundary between
different processes. If omitted the default value is 1. 
It's use will not be discussed further.

\end{itemize}

For the moment we will restrict ourselves to the study of lattices with the
basic (torus) topology, postponing the rules concerning the specifications
for {\tt mypartitioning} and {\tt mytopolgy}.

Here is a short program to create a 2D $8^2$ lattice (but no field
yet associated to it) running in parallel on 4 processes.
\begin{verbatim}
// Program: example14.C
#define PARALLEL
#define Nproc 4
#include "MDP_Lib2.h" 
#include "MDP_MPI.h" 
int main(int argc, char **argv) {
    mpi.open_wormholes(argc,argv);
    int mybox[]={8,8};
    generic_lattice mylattice(2,mybox);
    mpi.close_wormholes();
};
\end{verbatim} 

When it runs (since {\tt Nproc} is set to 4) the lattice 
will be automatically
subdivided into 4 sublattices of size \{2,8\}. {\tt mylattice}
contains information about the sites and how they are distributed among the
different processes. Moreover the different processes automatically
communicate to each other this information so that each of them knows which
local sites will have to be passed to the neighbor processes. All the
communications to establish the topology are done only once and the
information are inherited by all the fields defined on {\tt mylattice}.
The program produces the following output
\begin{verbatim}
PROCESS 0 STARTING
PROCESS 2 STARTING
PROCESS 1 STARTING
PROCESS 3 STARTING
=================================================================
Starting [ Matrix Distributed Processing ] ...
This program is using the packages: MDP_Lib2 and MDP_MPI
Created by Massimo Di Pierro (mdp@FNAL.GOV) version 17-11-1999
=================================================================
Going parallel ... YES

Initializing a generic_lattice...
Communicating...
Initializing random per site...
Done. Let's begin to work!
PROCESS 1 ENDING AFTER  0.010 sec.
PROCESS 2 ENDING AFTER  0.008 sec.
PROCESS 3 ENDING AFTER  0.011 sec.
=================================================================
Fractional time spent in communications by processor 0 is 0.03
Fractional time spent in communications by processor 1 is 0.04
Fractional time spent in communications by processor 2 is 0.02
Fractional time spent in communications by processor 3 is 0.03        
Ending this program.
Any runtime error below this point means mess with mpi and/or
deallocaton of memory.
=================================================================
PROCESS 0 ENDING AFTER  0.012 sec.
\end{verbatim} 

Note that the output automatically includes the execution time 
(using the wall clock) and the fractional time spent in 
communication by each processor (0.03 means 3\%). The latter
is computed by measuring the total time spent inside the 
communication function {\tt update()}.

Two other member functions of {\tt generic\_lattice} are
\begin{verbatim}
long generic_lattice::global_volume();
long generic_lattice::local_volume();
\end{verbatim} 
The first return the size of the total lattice volume (i.e. the total number
of sites), and the second returns the size of the portion of volume stored
by the calling process.

The member function
\begin{verbatim}
long generic_lattice::size();
\end{verbatim}
returns the total volume of the box containing the lattice and
\begin{verbatim}
long generic_lattice::size(int mu);
\end{verbatim} 
returns the size, in direction $\mu$, of the box containing the lattice.

\subsection{{\tt class site}}

\index{\tt site}
\index{\tt set()}
\index{\tt is\_in()}
\index{\tt is\_in\_boundary()}
\index{\tt is\_here()}
\index{\tt is\_equal()}
\index{\tt on\_which\_process()}
\index{\tt forallsites()}
\index{loop on sites}

To access a site one can define a {\tt site} variable using the syntax
\begin{verbatim}
site x(mylattice);
\end{verbatim} 

This tells the compiler that {\tt x} is a variable that will contain the
coordinate of a site and a reference to the lattice {\tt mylattice} on
which the point is defined.

Let's look at a modification of the preceding code that create a lattice
and, for each site {\tt x}, prints out the identification number of the
process that stores it:
\begin{verbatim}
// Program: example15.C
#define PARALLEL
#define Nproc 5
#include "MDP_Lib2.h" 
#include "MDP_MPI.h" 
int main(int argc, char **argv) {
    mpi.open_wormholes(argc,argv);
    int mybox[]={10,10};
    generic_lattice mylattice(2,mybox);
    site x(mylattice);
    forallsites(x)
       printf("Site=(%i,%i) is stored by %i\n", x(0), x(1), ME);
    mpi.close_wormholes();
};
\end{verbatim} 

Note that
\begin{verbatim}
x(mu)
\end{verbatim} 
returns the coordinate {\tt mu} of {\tt x}.

This simple program allows one to print a map of the lattice. The output of
this code depends on the number of processors on which it is running\footnote{%
The buffers of the different processors are copied to the standard output at
a random time partially scrambling the output of the different processes.
For this reason it is in general a good rule to print only from process {\tt %
ME==0}.}. The
loop command
\begin{verbatim}
forallsites(x) /* do something with x */
\end{verbatim} 
spans first all the even sites and then all the odd sites.

It is possible to set the value of a {\tt site} variable {\tt x} to a
particular site of given coordinates ($(3,7)$ for example) using the command
\begin{verbatim}
x.set(3,7);
\end{verbatim} 

\noindent but one must be careful and execute this statement only on the
process which contains the site $(3,7)$ or, eventually, in a process that
contains a neighbor of this site. In this case, in fact, the process will
also contains a copy of the original site therefore it will be able to
access the location. By default each process will store its own sites plus a
copy of next-neighbor sites in each direction and a copy of next-next
neighbor sites (moving in two different directions)\footnote{%
This is required, for example, by the clover term in QCD.}. 
They will be referred to as boundary sites. Boundary
sites are different for different processes.

It is possible to check if a site is a local site in the following way:
\begin{verbatim}
x.set(3,7)
if(x.is_in()) /* then do something */
\end{verbatim} 
\noindent The member function {\tt is\_in()} returns {\tt TRUE} if the site
is local. A better way to code it would be
\begin{verbatim}
if(on_which_process(3,7)==ME) {
   x.set(3,7);
   /* do something */
}
\end{verbatim} 
Other member functions of {\tt site} are
\begin{itemize}
\item {\tt is\_in\_boundary()} returns {\tt TRUE} if a site is in the
boundary of that process.

\item {\tt is\_here()} returns {\tt TRUE} is the site {\tt is\_in()} or {\tt %
is\_in\_boundary()}.

\item {\tt is\_equal(int,int,int...)} returns {\tt TRUE} 
if the site is equal to
the site specified by the coordinates that are arguments of {\tt is\_equal()}
(x0,x1,x2,...).{\bf \ Accessing a site that {\tt !is\_here()} crashes the
program.}

\end{itemize}

To move from one site to another is quite simple:
\begin{verbatim}
x=x+mu; // moves x up in direction mu (integer) of one step
x=x-mu; // moves x down in direction mu (integer) of one step
\end{verbatim} 

Note that {\tt mu} is an integer and in general

\begin{equation}
{\tt x+mu}\neq {\tt x}\neq {\tt x-mu} 
\end{equation}

even when {\tt mu} is zero.

Here is a test program to explore the topology:
\begin{verbatim}
// Program: example16.C
#define PARALLEL
#define Nproc 2
#include "MDP_Lib2.h" 
#include "MDP_MPI.h" 
int main(int argc, char **argv) {
    mpi.open_wormholes(argc,argv);
    int mybox[]={10,10};
    generic_lattice mylattice(2,mybox);
    site x(mylattice);
    int mu=0;
    if(ME==0) {
       x.set(0,0);
       do {
          printf("x=(%i,%i)\n", x(0), x(1));
         if(x.is_in_boundary()) error("I found the boundary");
         x=x+mu;
       } while(TRUE);
    };
    mpi.close_wormholes();
};
\end{verbatim} 

This program stops after 5 iterations if {\tt Nproc==2}, after 4 if {\tt %
Nproc==3}, after 3 if {\tt Nproc==4}, after 2 if {\tt Nproc==5} and so on.
It will never stop on a single process job ({\tt Nproc==1}) because of
periodic boundary conditions. Since all sites with the same {\tt x(0)}, by
default, are stored in the same process, if one moves in direction {\tt mu=1}
the job will never stop despite the number of processes.

\subsection{\tt class generic\_field}

\index{field variables}
\index{\tt generic\_field}
\index{\tt forallsites()}
\index{\tt forallsitesofparity()}
\index{\tt forallsitesandcopies()}
\index{loop on sites}
\index{\tt parity()}
\index{\tt EVEN}
\index{\tt ODD}

A {\tt generic\_field} is the simplest field one can define on a given {\tt %
generic\_lattice}.

Suppose, for example, one wants to associate a given structure, {\tt mystruct%
}, to each site of {\tt mylattice} (in the example a structure that contains
just a{\tt \ float}) and initialize the field variable to zero:
\begin{verbatim}
// Program: example17.C
#define PARALLEL
#define Nproc 2
#include "MDP_Lib2.h"
#include "MDP_MPI.h" 
struct mystruct { 
    float value; /* or any other structure */
};
int main(int argc, char **argv) {
    mpi.open_wormholes(argc,argv);
    int mybox[]={10,10};
    generic_lattice         mylattice(2,mybox);
    generic_field<mystruct> myfield(mylattice);
    site x(mylattice);
    forallsites(x)
      myfield(x).value=0;
    myfield.update();
    mpi.close_wormholes();
};
\end{verbatim} 

The command
\begin{verbatim}
generic_field<mystruct> myfield(mylattice);
\end{verbatim} 

\noindent defines {\tt myfield}, a field of {\tt mystruct} on the sites of 
{\tt mylattice.} The command {\tt myfield(x)}returns by reference the
structure associated to the site {\tt x}

The function
\begin{verbatim}
void generic_lattice::update(int,int,int)
\end{verbatim} 

\noindent is the most important of all. For the moment we will restrict to
that case when it is called with no arguments. Its task is to perform all the
communications necessary in order for each process to get an 
updated copy of the boundary sites stored by a neighbor process.

After the loop all sites that are local ({\tt forallsites}) are initialized,
but sites that are in boundary (since they are copies of sites initialized
on a different process) still contain ``random'' numbers. 
One way to initialize also the sites in
the boundary and avoid some time consuming communications is replacing the
lines
\begin{verbatim}
    forallsites(x)
      myfield(x).value=0;
    mylattice.update();
\end{verbatim} 

\noindent with
\begin{verbatim}
    forallsitesandcopies(x) myfield(x).value=0;
\end{verbatim} 

(Note that the command {\tt forallsitesandcopies} only works for a local
expression that does not call the local random generator).

It is also possible to loop on all the sites of a given parity, for
example EVEN (or ODD), in a slow way
\begin{verbatim}
    int parity=EVEN;
    forallsites(x) 
      if(x.parity()=parity)
        /* then do something */
\end{verbatim} 

\noindent or in a fast way

\begin{verbatim}
    int parity=EVEN;
    forallsitesofparity(x,parity)
      /* do something */ 
\end{verbatim} 

\noindent The second expression is faster because sites with the same 
parity are stored contiguously in memory.

\subsubsection{Local random generator}

\index{\tt random\_generator\_class}
\index{local random generator}

Once a {\tt %
generic\_lattice} is declared one random generator per site is
automatically created and initialized with a function of the site
coordinates. The random generator (and its member functions) of site {\tt x}
of {\tt mylattice} can be accessed (by reference) with the command
\begin{verbatim}
Random_generator& generic_lattice::random(site) 
\end{verbatim} 

Program {\tt example17.C} can be modified 
so that {\tt myfield} is initialized, for
example, with a gaussian random number:
\begin{verbatim}
// Program: example18.C
#define PARALLEL
#define Nproc 2
#include "MDP_Lib2.h"
#include "MDP_MPI.h" 
struct mystruct {
    float value; /* or any other structure */
};
int main(int argc, char **argv) {
    mpi.open_wormholes(argc,argv);
    int mybox[]={10,10};
    generic_lattice         mylattice(2,mybox);
    generic_field<mystruct> myfield(mylattice);
    site x(mylattice);
    forallsites(x)
      myfield(x).value=mylattice.random(x).gaussian();
    myfield.update();
    mpi.close_wormholes();
};
\end{verbatim} 

\subsubsection{Accessing a lattice from a field}

\index{\tt lattice()}

It is always possible to access a {\tt generic\_lattice} from its field
using {\tt lattice()}, a member function of {\tt %
generic\_field} that returns by reference the {\tt generic\_lattice} on
which the field is defined.

For example in the preceding program one could substitute
\begin{verbatim}
myfield(x).value=mylattice.random(x).gaussian();
\end{verbatim} 

\noindent with
\begin{verbatim}
myfield(x).value=myfield.lattice().random(x).gaussian();
\end{verbatim} 

\noindent and get the same result. This is useful because one can
avoid passing a {\tt generic\_lattice} object to a function of 
a {\tt generic\_field}.

\subsubsection{More on {\tt update()}}

\index{\tt update()}

In some cases it may be useful to have a field that contains {\tt n
mystruct} variables at each site. 

A field of this type can be defined with tyhe command
\begin{verbatim}
generic_field<mystruct> myfield(mylattice,n);
\end{verbatim} 

Consider, for example, the simple case of a tensor field 
$T_\mu ^{ij}(x),$ where $%
i,j\in \{0..5\}$, $\mu \in \{0...3\}$. It can be defined with
\begin{verbatim}
generic_field<Complex[4][5][5]> T(mylattice);
\end{verbatim} 

\noindent and its elements can be accessed with the natural expression:
\begin{verbatim}
T(x)[mu][i][j] 
\end{verbatim} 
where {\tt mu,i,j} are integers.

Alternatively the tensor $T_\mu ^{ij}$ can be defined with the command
\begin{verbatim}
generic_field<Complex[5][5]> T(mylattice,4);
\end{verbatim} 

\noindent and its elements can be accessed with the expression:
\begin{verbatim}
T(x,mu)[i][j]
\end{verbatim} 

Sometimes the second choice is better for the following reason: 
In many algorithms one often needs
to update all the boundary site variables having a given a parity and a
given {\tt mu}. One wants to do the update in one single communications.
The {\tt update()} function allows one to do exactly this:
\begin{verbatim}
int parity=EVEN, int mu=3;
T.update(parity,mu);
\end{verbatim} 

\subsubsection{{\tt operator() }}

\index{field of \tt Matrix}

Some care is required when declaring a {\tt generic\_field}. The structure
associated to the sites cannot contain a pointer to dynamically allocated
memory, therefore it cannot contain a {\tt Matrix} (or a {tt DynamicArray})
object. This is not a
limitation since the structure associated to the sites can be a
multidimensional array and one can always map it into a {\tt Matrix} by
creating a new field, inheriting {\tt generic\_field} and redesigning {\tt %
operator()}. 

In this subsection we explain how to use {\tt Matrix} to build and 
interface to the field variables.

Let's go back to the example of a tensor field $T_{\mu }^{ij}(x)$. One can
define it as
\begin{verbatim}
class mytensor: public generic_lattice<Complex[5][5]> {};
mytensor T(mylattice,4);
\end{verbatim} 

Instead of accessing its elements as {\tt Complex} numbers one may want to
access the two-dimensional array as a {\tt Matrix} object. This is done by
overloading the {\tt operator()} of the class generic\_field:
\begin{verbatim}
Matrix myfield::operator() (site x, int mu) {
   return Matrix(address(x,mu),5,5);
};
\end{verbatim} 

\noindent where {\tt address()} is a member function of {\tt %
generic\_lattice} that returns the location of memory where the field
variables at {\tt (x,mu)} are stored.
After this definitions one can simply access the field
using the expression:
\begin{verbatim}
T(x,mu)
\end{verbatim} 
that now returns a {\tt Matrix}.

Now {\tt mytensor T} looks like a field of $5\times 5$ matrices and its
elements can be used as discussed in chapter 2. For example one can
initialize the tensor using random $SU(5)$ matrices generated
independently by the local random generators associated to each site
\begin{verbatim}
forallsites(x)
  for(mu=0; mu<4; mu++)
    T(x,mu)=T.lattice().random(x).SU(5);
\end{verbatim} 

\noindent or print a particular element
\begin{verbatim}
x.set(3,5);
if(x.is_in()) print(T(x,2));
\end{verbatim} 

\subsubsection{A few derived fields}

Following the directives of the last subsections four more basic
fields are implemented in {\tt MDP 1.2}.
\begin{itemize}

\item {\tt Matrix\_field}: a field of {\tt Matrix};
\item {\tt NMatrix\_field}: a field of n {\tt Matrix};
\item {\tt Vector\_field}: a field of vectors (a vector
is seen as a {\tt Matrix} with one single column);
\item {\tt NVector\_field}: a field of n vectors;
\end{itemize}

To explain their usage we present here a few lines of 
code that define a field of $3 \times 4$ matrices, {\tt A}, a field of
4-vectors, {\tt u}, and a field of 3-vectors, {\tt v}; then for each
site compute 
\begin{equation}
v(x)=A(x) u(x)
\end{equation}
This can be implemented as
\begin{verbatim}
Matrix_field A(mylattice,3,4);
Vector_field u(mylattice,4);
Vector_field v(mylattice,3);
/* assign values to the fields */
forallsites(x) v(x)=A(x)*u(x);
\end{verbatim} 
and its generalization to 
\begin{equation}
v_i(x)=A_i(x) u_i(x)
\end{equation}
for {\tt i} in the range [0,N-1]
\begin{verbatim}
int N=10; // user defined variable
NMatrix_field A(mylattice,N,3,4);
NVector_field u(mylattice,N,4);
NVector_field v(mylattice,N,3);
/* assign values to the fields */
forallsites(x) for(i=0; i<N; i++) v(x,i)=A(x,i)*u(x,i);
\end{verbatim} 

\subsection{Input/Output}

\index{Input}
\index{Output}
\index{\tt save()}
\index{\tt load()}

The class {\tt generic\_field} contains two I/O member functions:
\begin{verbatim}
void generic_lattice::save(char[],int,int);
void generic_lattice::load(char[],int,int);
\end{verbatim} 

They take three arguments

\begin{itemize}
\item  The{\tt \ filename} of the file to/from which to save/load the data

\item  The identification number of the master process that is will 
physically perform to the I/O

\item  The size (in sites) of the buffer used by the master process to
communicate with the other processes.
\end{itemize}

The last two arguments are optional and by default the master process is
process 0 and the buffer size is 1024 sites. The identification number of
the master process has to correspond to the process running on the processor
that is physically connected with the I/O device where the file {\tt filename%
} is. The size of the buffer will not affect the result of the I/O operation
but may affect the speed (which should increase with the size of the
buffer). If the size is too big there may be an ``out of memory'' problem
(that usually results in obscure error messages when compiling with
MPI). Note that the I/O functions
save/load have to be called by all processes (not just the master one). In
fact all processes which contain sites of the lattice are involved in
the I/O by exchanging informations with the master.

As an example of I/O we consider the case of a {\tt save()} operation. Each
of the processes arranges the local field variables to be saved into
packets and sends the packets to the process that performs the I/O (we
assume it is process {\tt 0}).
Process {\tt i} arranges its site into packets ordered according with
the global parameterization and sends one packet at the time to process
{\tt 0}. In this way process {\tt 0} only receives one packet at the time
for each of the processes and already finds the sites stored in the
correct order so that it does not need to perform a seek. Process
{\tt 0} saves the sites according to the global 
parameterization (which is
independent from the lattice partitioning over the parallel processes).
Its only task is to sweep all the sites and, for each of them, 
pick the corresponding field variables from the packet sent by the 
process that stored it.

We believe this is the most efficient way to implement a parallel I/O
for field variables. As a particular case, assuming process {\tt 0}
has enough memory, one could set the size of the buffer equal to the
maximum number of sites stored by a process. In this case the parallel
I/O would be done performing only one communication per process.

As an example of efficiency: 
a field as large as 100MB can be read and distributed
over 8 500MHz PentiumIII PCs in parallel (connected with Ethernet) 
in a few of seconds.

The save/load function are inherited by every field based on {\tt %
generic\_field}. For example:
\begin{verbatim}
/* define mylattice */
class myfield: public generic_field<Complex[7][3]> {};
myfield F(mylattice);
/* do something */
F.save("test.dat");
F.load("test.dat");
\end{verbatim} 

If a field contains member variables other than the structure at
the sites, these member variables are ignored by save/load. The only
variables that are saved together with the field are:
\begin{itemize}
\item the number of dimensions
\item the size of the box containing the lattice
\item the total number of sites belonging to the lattice 
\item the size in bytes of the structure at each lattice site
\item a code to detect the endianess of the computer that saved the
data
\item the time and date when the file was created
\end{itemize}

We also provide a small program, {\tt inspect.C}, that can open a
file and extract this information.

It is always possible to add any other information to the bottom of the
file (in binary or ASCII format) without compromising its integrity.

\subsection{The internal representation of a lattice }

\index{coordinates!local}
\index{coordinates!global}
\index{partitioning}
\index{topology}

\begin{figure}
\epsfxsize=7cm
\epsfysize=7cm
\hfil \epsfbox{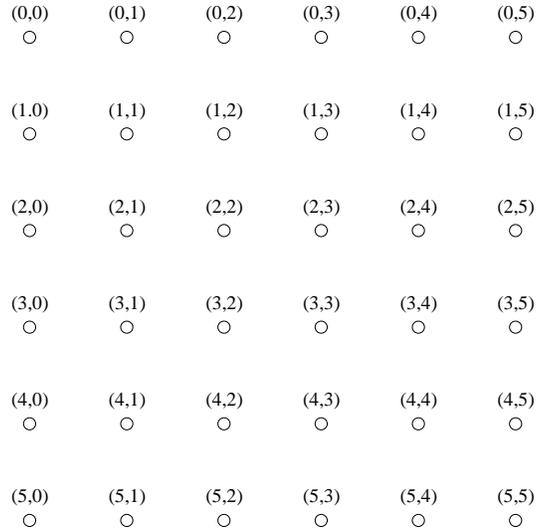} \hfil
\caption{Example of a $6 \times 6$ grid. The points are labelled by their coordinates.} 
\end{figure}

\begin{figure}
\epsfxsize=7cm
\epsfysize=7cm
\hfil \epsfbox{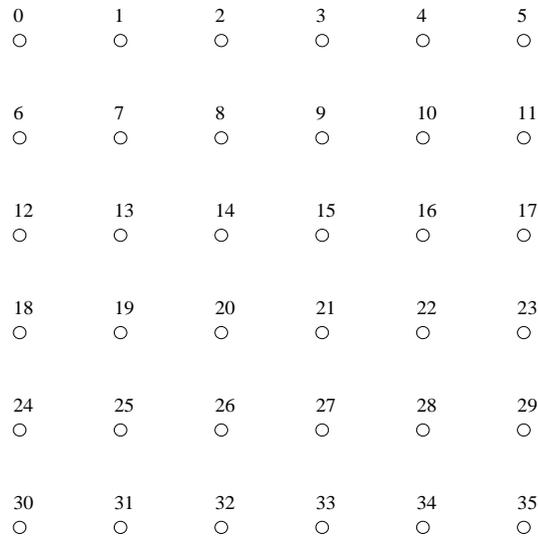} \hfil
\caption{Example of a $6 \times 6$ grid. The points are labelled by their global index.}
\end{figure}

A lattice is a collection of sites and, for sake of simplicity, from now on
we will refer to the lattice defined by
\begin{verbatim}
int ndim=2;
int mybox[]={6,6};
int mypartitioning(int x[], int ndim, int nx[]) {
   if(x[0]<3) return 0;
   if(x[1]<4) return 1;
   return 2;
};
generic_lattice mylattice(ndim,mybox,mypartitioning); 
\end{verbatim} 

\noindent This lattice is represented in fig.2 and partitioned as in fig.5.

To speed up the communication process {\tt MDP 1.2} uses different
parametrizations for labelling the sites:

\begin{itemize}
\item  the global coordinates (as shown in fig.2)

\item  a global index (as shown in fig.3)

\item  a local index (as shown in fig.8 from the point of view of process 0)
\end{itemize}

\noindent and it provides functions to convert one parameterization 
into another.

\begin{figure}
\epsfxsize=7cm
\epsfysize=7cm
\hfil \epsfbox{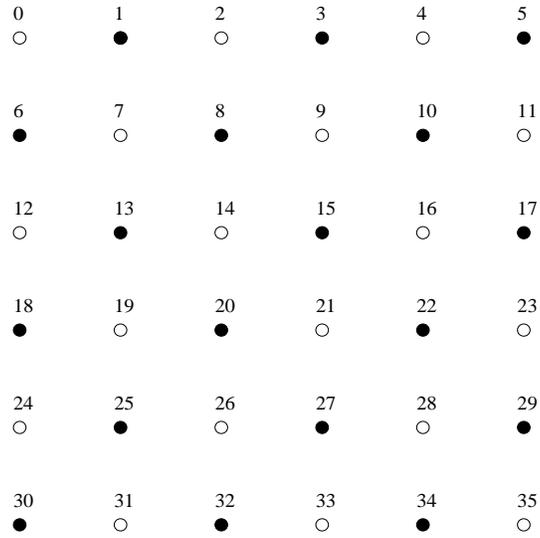} \hfil
\caption{Example of a $6 \times 6$ grid. The points are labelled by their global index. Points with even (white) and odd (black) parity are distinguished.}\end{figure}

\begin{figure}
\epsfxsize=7cm
\epsfysize=7cm
\hfil \epsfbox{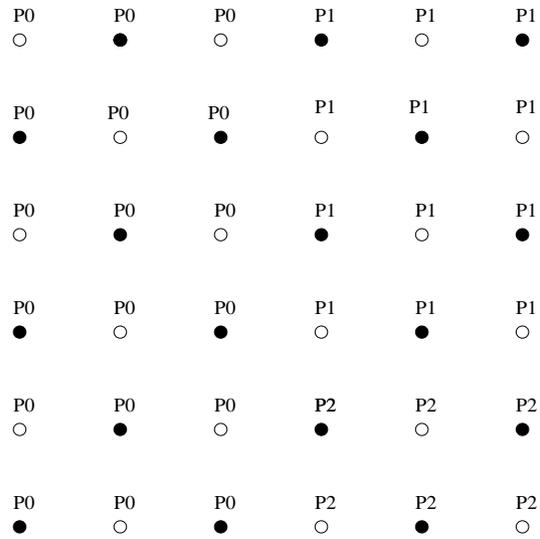} \hfil
\caption{Example of a $6 \times 6$ grid distributed on 3 processors (according with some user defined partitioning). Each point is labelled by the identification number of the process that stores it.}
\end{figure}

The global parameterization is quite natural. For example: the point {\tt %
x[]=\{3,2\}} has a global index given by {\tt x[1]*mybox[0]+x[0]}; The two
global parametrizations have a one to one correspondence. The local
parameterization is more complicated and it represents the order in which
each process stores the sites in the memory.

When the lattice is defined the constructor of the class {\tt %
generic\_lattice} spans all the sites using the global coordinates (fig.2)
and it assigns them the global index (fig.3) and a parity (fig.4). Then each
process checks which sites are local sites (fig.5). After the topology is
assigned to the lattice (fig.6), each process identifies which sites are
neighbors of the local sites (fig.7 for process 0). At this point each
process has all the information needed to build a local parameterization for
the sites. All even sites belonging to process 0 are labeled with a
progressive number; then all the odd sites from process 0 are labelled in
progression. The same procedure is repeated by each process for each process
(including itself). The result (for the case in the example) is shown in
fig.8 (from the point of view of process 0).

When a field is associated to the lattice the field variables are stored by
each process according with the local parameterization. This guarantees that:

\begin{figure}
\epsfxsize=7cm
\epsfysize=7cm
\hfil \epsfbox{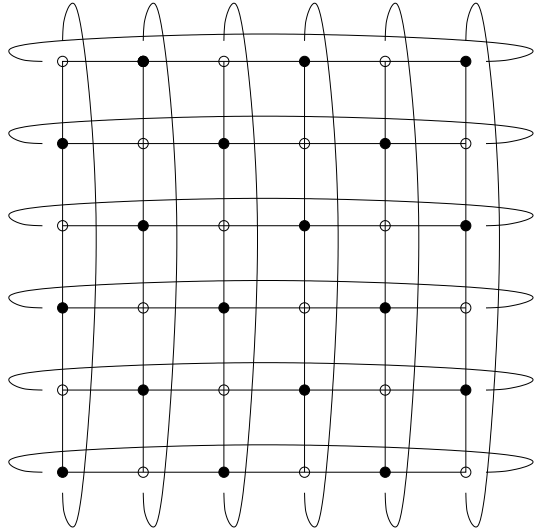} \hfil
\caption{Example of a $6 \times 6$ grid endowed with the default (torus) topology. When the topology is assigned to the grid, it is promoted to the rank of lattice.}
\end{figure}

\begin{figure}
\epsfxsize=7cm
\epsfysize=7cm
\hfil \epsfbox{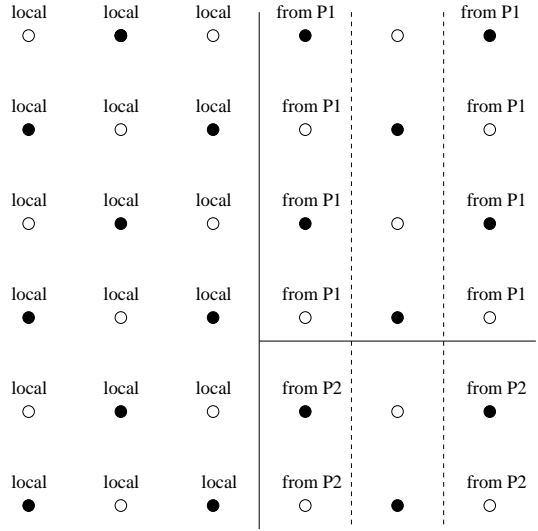} \hfil
\caption{Example of a $6 \times 6$ lattice form the point of view of process 0.
Process 0 knows which sites are local, which non-local sites are neighbors of the local sites and from which process to copy them.}
\end{figure}

\begin{figure}
\epsfxsize=7cm
\epsfysize=7cm
\hfil \epsfbox{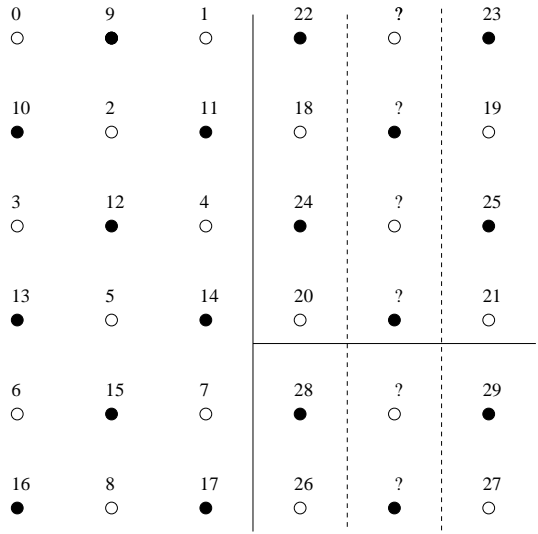} \hfil
\caption{Example of a $6 \times 6$ lattice form the point of view of process 0. Process zero assigns a local parameterization to the local sites and to the neighbor sites of the local points. It cannot access the rest of the sites.}
\end{figure}

\begin{itemize}
\item  Boundary sites that have to be copied from the same process are
stored continuously in the memory. This minimizes the need for communications.

\item  Boundary sites copied from the same process (and the local sites as
well) of given parity are also stored continuously in memory. This speeds up
loops on sites of given parity.

\item  When a process X copies a neighbor site from process Y it does not
 need to allocate a buffer because the data can be received directly in
the memory location where it is normally stored.
\end{itemize}

\index{\tt global\_index()}
\index{\tt local\_index()}
\index{\tt set\_global()}
\index{\tt set\_local()}

Given a site variable {\tt x} on {\tt mylattice} one can ask for the {\tt mu}
coordinate using the sintax

{\tt x(mu)}

\noindent or the global index

{\tt x.global\_index()}

\noindent or the local index

{\tt x.local\_index()}

\noindent Moreover one can initialize a site variable specifying the coordinates

{\tt x.set(3,2);}

\noindent the global index

{\tt x.set\_global(26);}

\noindent or the local index

{\tt x.set\_local(7); /* assuming ME is 0 */}

\noindent If {\tt x} and {\tt y} are two {\tt site} variables in the
same code defined  on the same lattice, but using different
partitioning. One can assign y to x using the assignment operator

{\tt x=y;}

\noindent This is equivalent to

{\tt x.set\_global(y.global\_index());}

\subsection{Partitioning and topology }

\index{coordinates!local}
\index{coordinates!global}
\index{partitioning!define}
\index{topology!define}

One of the most important characteristics of {\tt MDP 1.2} is
that one is not limited to a box-like lattice (even if it has to be a
subset of the points in a box) and the sites of the lattice can be
associated to an arbitrary topology. Before stating with weird staff
here is the definition of the default function for the lattice
partitioning:
\begin{verbatim}
template<int dim>
int default_partitioning(int *x, int ndim, int *nx) { 
  float tpp1=nx[dim]/NPROC; 
  int   tpp2=(int) tpp1; 
  if(tpp1-tpp2>0) tpp2+=1;
  return x[dim]/tpp2; 
};
\end{verbatim} 

Any partitioning function takes three arguments:

\begin{itemize}
\item  An array {\tt x[ndim]} containing the ndim coordinates of a site

\item  The dimensions, {\tt ndim}, of the box containing the lattice
sites

\item  An array, {\tt nx[ndim],}containing the size of box in each
dimension
\end{itemize}

It returns the identification number of the process that stores the
site corresponding to the coordinates in {\tt x[]}\footnote{
that at this level {\tt site} variables are not used. In fact the
partitioning function (as well the topology function) is called before
the lattice is defined, and it is used to define it. A site variable
can be defined only after a lattice is defined.}.

One can define a different partitioning just by defining another
function that takes the same arguments as the default one
and pass it as third argument
to the constructor of {\tt generic\_lattice}. If the new partitioning
function, for a particular input site, returns a number outside the
range {\tt 0 to NPROC-1,} that site is excluded from the lattice. To
this scope the macro  {\tt NOWEHERE }is defined. If a site has to be
excluded from the lattice the partition function should return {\tt
NOWHERE}. If one defines a partitioning function that excludes some
points from the lattice one is forced to create e user defined
topology, otherwise one ends up with sites which are neighbors of
sites that are {\tt NOWHERE} and the program fails.

The default torus topology is defined as
\begin{verbatim}
void torus_topology(int mu, int *x_dw, int *x, int *x_up, 
                    int ndim, int *nx) { 
  for(int nu=0; nu<ndim; nu++) if(nu==mu) {
    x_dw[mu]=(x[mu]-1+nx[mu]) % nx[mu]; 
    x_up[mu]=(x[mu]+1) % nx[mu]; 
  } else x_up[nu]=x_dw[nu]=x[nu]; 
};
\end{verbatim} 

Any topology function takes six arguments, they are:

\begin{itemize}
\item  A direction {\tt mu} in the range {\tt 0 to ndim-1}

\item  A first array of coordinates, {\tt x\_dw[]}

\item  A second array of coordinates, {\tt x[]}

\item  A third array of coordinates, {\tt x\_up[]}

\item  The dimensions, {\tt ndim}, of the box containing the lattice
sites

\item  An array, {\tt nx[ndim],}containing the size of box in each
dimension
\end{itemize}

The coordinates in {\tt x[]} are taken as input and the topology
function fills the the arrays {\tt x\_dw[]} and {\tt x\_up[]} with the
coordinates of the neighbor points when moving from {\tt x[]} one step
in direction {\tt mu} down and up respectively.

The topology function will only be called for those sites {\tt x[]}
that belong to the lattice. It has to fill {\tt x\_dw[]} and {\tt
x\_up[]} with the coordinates of points that have not been excluded
from the lattice.

\subsubsection{Non periodic lattice}

\index{lattice!finite}

As an example one can create a lattice with a true physical boundary
by saying, for example, that moving up (or down) in a particular
direction from a particular subset of sites one remains where one
is\footnote{ This kind of lattice can be used, for example, to study
a 3D solid under stress and study its deformations.}. In the simple
case of a finite box one could define the topology:
\begin{verbatim}
void box_topology(int mu, int *x_dw, int *x, int *x_up,  
                  int ndim, int *nx) {  
  torus_topology(mu,x_dw,x,x_up,ndim,nx); 
  if(x[mu]==0) x_dw[mu]=x[mu]; 
  if(x[mu]==nx[mu]-1)) x_up[mu]=x[mu]; 
};
\end{verbatim} 

\subsubsection{Hexagonal lattice}

\index{lattice!exagonal}

\begin{figure}
\epsfxsize=10cm
\epsfysize=5cm
\hfil \epsfbox{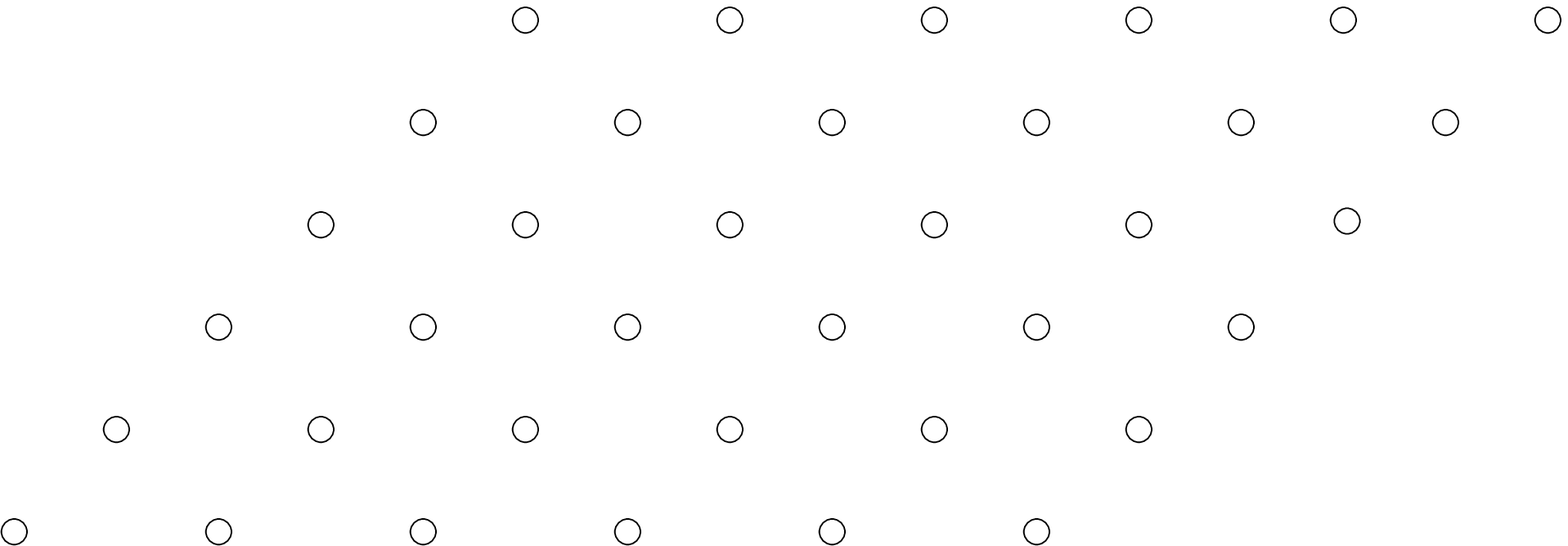} \hfil
\vskip 1cm
\epsfxsize=10cm
\epsfysize=5cm\hfil \epsfbox{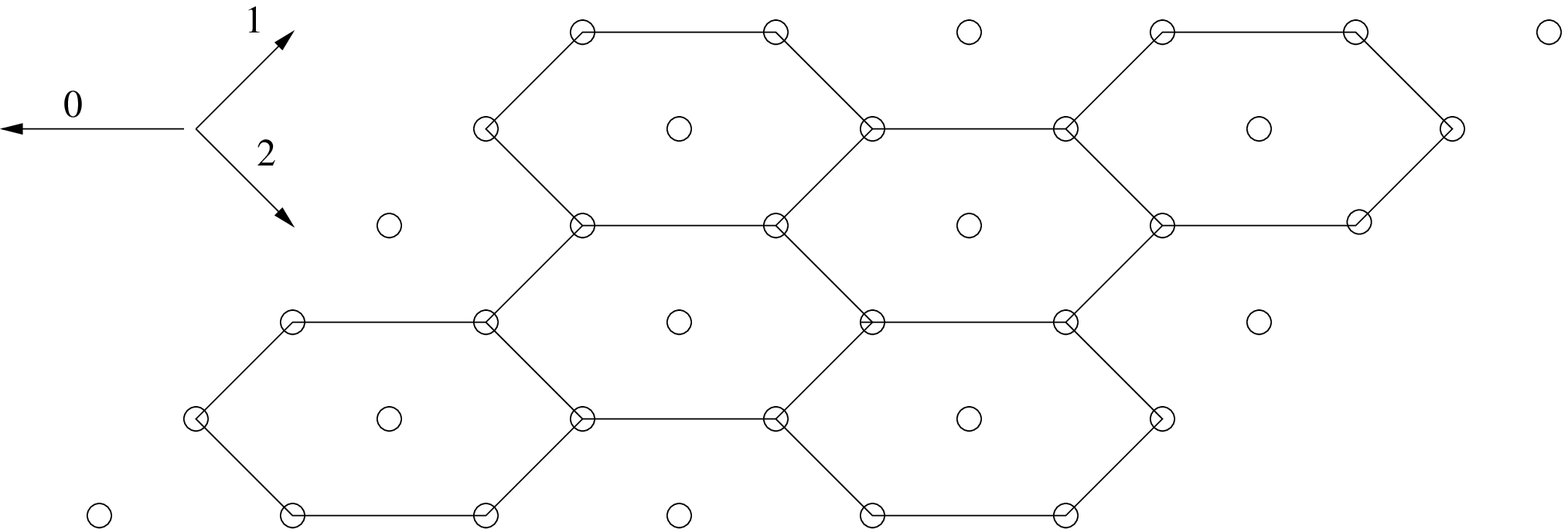} \hfil
\caption{Example of a $6 \times 6$ grid to be used to build an hexagonal lattice (top) and the metric associated to this grid (bottom). the points at center of the hexagons are excluded from the lattice by the partitioning function.}
\end{figure}

Another example of a lattice with a weird topology is the 
hexagonal grid shown in fig.9.  
The trick to do it is having a 3D lattice flat in the third
direction, put NOWHERE the centers of the hexagons and define the
proper topology for the remaining points. It can be done by the
following code
\begin{verbatim}
int exagonal_partitioning(int x[], int ndim, int nx[]) {
  if((x[0]+x[1])%3==0) return NOWHERE; 
  else return default_partitioning<0>(x,ndim,nx); 
};
 
void exagonal_topology(int mu, int *x_dw, int *x, int *x_up, 
                       int ndim, int *nx) { 
  if((x[0]+x[1])%3==1) {
    switch(mu) { 
    case 0: x_up[0]=(x[0]+1)%nx[0]; x_up[1]=x[1]; break; 
    case 1: x_up[1]=(x[1]+1)%nx[1]; x_up[0]=x[0]; break; 
    case 3: x_up[0]=(x[0]-1+nx[0])%nx[0]; 
            x_up[1]=(x[1]-1+nx[1])%nx[1]; break; 
    };
    x_up[2]=x[2]; x_dw[0]=x[0]; 
    x_dw[1]=x[1]; x_dw[2]=x[2]; 
  };
  if((x[0]+x[1])%3==2) { 
    switch(mu) { 
    case 0: x_dw[0]=(x[0]-1+nx[0])%nx[0]; x_dw[1]=x[1]; break; 
    case 1: x_dw[1]=(x[1]-1+nx[0])%nx[1]; x_dw[0]=x[0]; break; 
    case 3: x_dw[0]=(x[0]+1)%nx[0]; x_dw[1]=(x[1]+1)%nx[1]; break; 
    };
    x_dw[2]=x[2]; x_up[0]=x[0]; 
    x_up[1]=x[1]; x_up[2]=x[2]; 
  }; 
}; 

int n=2;
int mybox[]={3*n, 3*n, 1}; 

generic_lattice exagonal(3,mybox,exagonal_partitioning,  
                         exagonal_topology);
\end{verbatim} 

One can observe that we need a flat third dimension to allow for
three possible direction. For this reason {\tt MDP 1.2} allows to
define a lattice with a number of directions different from the number
of dimensions. To avoid entering in a number of technical details we
will avoid discussing this possibility here.
What one can do with such a lattice is another matter and will
not be discussed. From now on we will concentrate on more regular lattices
which tend to have a more intuitive interpretation.

\subsubsection{Antiperiodic boundary conditions}

\index{lattice!aperiodic}

There are two possible meaning for antiperiodic boundary conditions:
1)   antiperiodic boundary conditions for the lattice (i.e. a Moebious
topology).  2) antiperiodic boundary conditions for a field defined on
the lattice.

The first case can be handled in the same way as any other
topology. The second case has nothing to do with the lattice topology,
but is a property of the field and one has to take care of
it in the definition of {\tt operator()} for the field in question.

\subsection{Memory optimization }

\index{memory usage}
\index{\tt deallocate\_memory()}
\index{\tt allocate\_field()}

Actually the class {\tt generic\_field} has more properties of those
described. For example each field can be deallocated at any time if
memory is needed:
\begin{verbatim}
struct mystruct { /* any structure */ }; 
generic_lattice mylattice1(mydim1,mybox1); 
generic_field<mystruct> myfield(mylattice1);
/* use myfield */ 
myfield.deallocate_memory();  
generic_lattice mylattice2(mydim2,mybox2); 
myfield.allocate_field(mylattice2);
\end{verbatim} 
Note that one cannot reallocate a field using a different structure at
the site. This would be very confusing.

\section{Examples of parallel applications}

We will present here a few example of parallel application that use
{\tt MDP 1.2}. To keep things simple we will not make use of fields of
matrices.

\subsection{Electrostatic potential for a given distribution of charges}

\index{electrostatic potential}

We consider here the vacuum space inside
a cubic metal box connected to ground and containing a given (static)
distribution of charge. We want to determine the electrostatic
potential in the box.

The vacuum is implemented as a finite $20^3$ lattice. Two fields,
a charge density $q(x)$ and a potential $u(x),$ are defined on
it. In the continuum the potential is obtained by solving the Gauss
law equation
\begin{equation}
\triangledown ^2u(x)=q(x)
\end{equation}
It discretized form reads
\begin{equation}
\sum_{i=0}^2\left[ u(x+\widehat{i})-2u(x)+u(x-\widehat{i})\right] =q(x)
\end{equation}
which can be solved in $u(x)$%
\begin{equation}
u(x)=\frac 16\left[ q(x)+\sum_{i=0}^2u(x+\widehat{i})+u(x-\widehat{i}%
)\right]   \label{gauss3}
\end{equation}
Therefore the static potential solution is obtained by iterating
eq.(\ref{gauss3}) on the vacuum until convergence.

As initial condition we assume that only two charges are present
in the box:
\begin{equation}
q(x)=3\delta (x-A)-5\delta (x-B)
\end{equation}
where $A=(3,3,3)$ and $B=(17,17,17)$.  Moreover since the box has a
finite extension the {\tt box\_topology} will be used.

Here is the parallel program based on {\tt MDP 1.2}.

\begin{verbatim}
// Program: application1.C 
#define PARALLEL
#define Nproc 4
#include "MDP_Lib2.h" 
#include "MDP_MPI.h" 
int main(int argc, char **argv) { 
    mpi.open_wormholes(argc,argv);
    int mybox[]={20,20,20}; 
    generic_lattice vacuum(3,mybox,
                           default_partitioning<0>,
                           box_topology);
    generic_field<float> u(vacuum); 
    generic_field<float> q(vacuum);
    site x(vacuum);
    site A(vacuum);
    site B(vacuum);
    A.set(3,3,3);
    B.set(17,17,17);
    float precision, old_u;
    forallsitesandcopies(x) { 
      u(x)=0;
      if(x==A)      q(x)=3;
      else if(x==B) q(x)=-5;
      else          q(x)=0;
    };
    do {
      precision=0;
      forallsites(x) if((x(0)>0) && (x(0)<mybox[0]-1) &&
                        (x(1)>0) && (x(1)<mybox[1]-1) &&
                        (x(2)>0) && (x(2)<mybox[2]-1)) {
        old_u=u(x);
        u(x)=(q(x)+u(x+0)+u(x-0)+u(x+1)+u(x-1)+u(x+2)+u(x-2))/6;
        precision+=pow(u(x)-old_u,2);
      };
      u.update();
      mpi.add(precision);
    } while (sqrt(precision)>0.0001);
    u.save("potential.dat");
    mpi.close_wormholes(); 
};
\end{verbatim} 

The saved values can be used to produce a density plot representing
the potential in the volume.

\subsection{Total impedance in net of resistors}

\index{net of resistors}
\index{impedence}

\begin{figure}
\epsfxsize=10cm
\epsfysize=7cm
\hfil \epsfbox{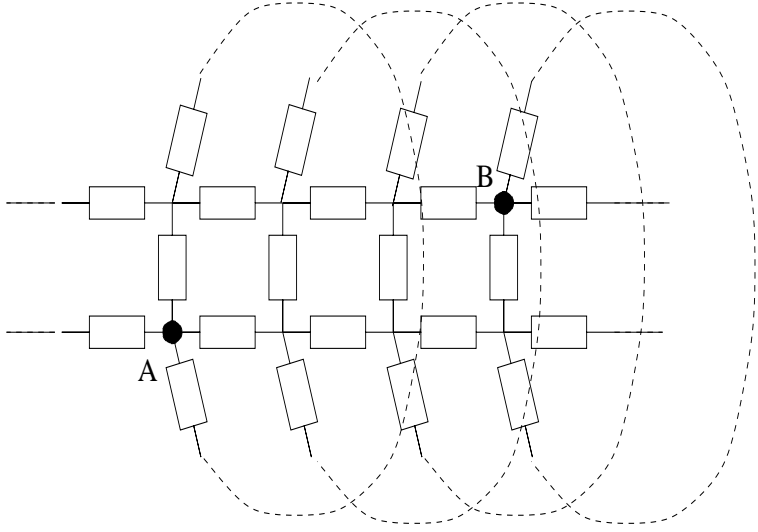} \hfil
\caption{Example of a cylindrical net of resistors.}
\label{resnet}
\end{figure}

Another problem we want to solve is that of determining the total
resistance between two arbitrary points (A and B) on a semi-conducting
finite cylindrical surface. The cylinder is obtained
starting from a torus topology and cutting the torus in one direction
(say 0). The total resistance is probed by connecting the terminals
of a current generator ($J$) to the points A and B and evaluating
the  potential at points A and B using Ohm's law
\begin{equation}
R_{AB}=\frac{u(A)-u(B)}J
\end{equation}
The surface of the semiconductor is implemented as a periodic net
of resistances so that each site has four resistances connected to
it, fig.~\ref{resnet}. 
For simplicity we assume that all the resistances are the
same (i.e the material is homogeneous). A different choice would be
equally possible.

The potential $u(x)$ is computed using the Kirchoff law at each
vertex
\begin{equation}
\sum_{i=0,1}R_i(x-\widehat{i})\left[ u(x-\widehat{i})-u(x)\right]
+R_i(x)\left[ u(x+\widehat{i})-u(x)\right] +J=0
\end{equation}
that, solved in $u(x)$, leads to
\begin{equation}
u(x)=\frac{\sum_{i=0,1}R_i(x-\widehat{i})u(x-\widehat{i})+R_i(x)u(x+\widehat{%
i})+J}{\sum_{i=0,1}R_i(x-\widehat{i})+R_i(x)}
\end{equation}
This equation is iterated on each site until convergence.

Here is the parallel program based on {\tt MDP 1.2}.

\begin{verbatim}
// Program: application2.C 
#define PARALLEL
#define Nproc 4 
#include "MDP_Lib2.h"
#include "MDP_MPI.h" 
void open_cylinder(int mu, int *x_dw, int *x, int *x_up, 
                   int ndim, int *nx) {
   torus_topology(mu,x_dw,x,x_up,ndim,nx);
   if((mu==0) && (x[0]==0))     x_dw[0]=x[0];
   if((mu==0) && (x[0]==nx[0]-1)) x_up[0]=x[0];
};
float resistance(int x0, int x1, int mu) {
   return (Pi/100*x0);
};
int main(int argc, char **argv) { 
    mpi.open_wormholes(argc,argv);
    int mybox[]={100,20}; 
    generic_lattice cylinder(2,mybox,default_partitioning<0>,
                             open_cylinder); 
    generic_field<float> u(cylinder); 
    generic_field<float> r(cylinder,2);
    site x(cylinder);
    site A(cylinder);
    site B(cylinder);
    float precision, old_u;
    float c, J=1, deltaJ, deltaR, Rtot, local_Rtot;
    A.set(15,7);
    B.set(62,3);
    forallsitesandcopies(x) { 
      u(x)=0;
      r(x,0)=resistance(x(0),x(1),0);
      r(x,1)=resistance(x(0),x(1),1);
    };
    do {
      precision=0;
      forallsites(x) {
         old_u=u(x);
         if(x==A) { c=+J; printf("%i\n", ME); };
         if(x==B) c=-J;
         deltaJ=c; deltaR=0;
         /* the next two lines take care of the finite cylinder */
         if(x+0!=x) { deltaJ+=u(x+0)*r(x,0);   deltaR+=r(x,0); };
         if(x-0!=x) { deltaJ+=u(x-0)*r(x-0,0); deltaR+=r(x-0,0); };
         deltaJ+=u(x+1)*r(x,1);   deltaR+=r(x,1); 
         deltaJ+=u(x-1)*r(x-1,1); deltaR+=r(x-1,1); 
         u(x)=deltaJ/deltaR;
         precision+=pow(u(x)-old_u,2);
      };
      u.update();
      mpi.add(precision);
    } while (sqrt(precision)>0.00001);
    Rtot=0;
    if(A.is_in()) Rtot+=u(A)/J;
    if(B.is_in()) Rtot-=u(B)/J;
    mpi.add(Rtot);
    if(ME==0) printf("R_A-R_B=%f\n", Rtot);
    mpi.close_wormholes(); 
};
\end{verbatim} 

\subsection{Ising model}

\index{Ising model}

\begin{figure}
\epsfxsize=8cm
\epsfysize=6cm
\hfil \epsfbox{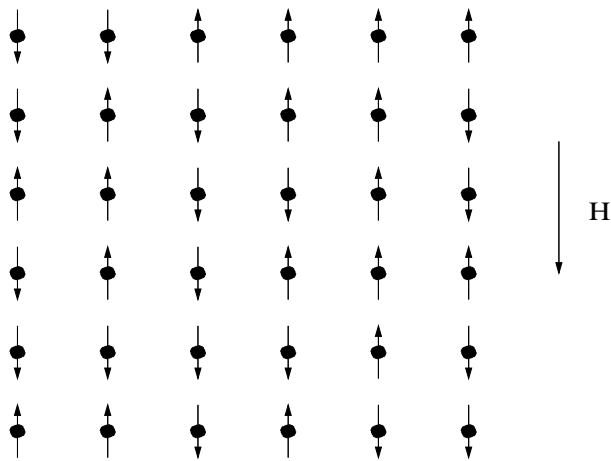} \hfil
\caption{Example of a 2D spin system.}
\label{spinlattice}
\end{figure}

One more full application we wish to consider is the study of the
total magnetization of a spin system under a magnetic field (this is
also known as the Ising model) as function of the temperature. 
As usual we define a lattice that represents the geometry of the
spin system, and
associate to the sites a field $s(x) \in \{-1, +1\}$. This is schematically represented in fig.~\ref{spinlattice}. The Hamiltonian of this system is
\begin{equation}
H=-\beta \sum_{x,y} s(x) \Delta(x,y) s(y) - \kappa \sum_x s(x) h(x) 
\end{equation}
\noindent where $h(x)$ is an external 
magnetic field, $\beta=1/T$ is the inverse
temperature and $\kappa$ is a constant typical of the material (in
principle one could promote $\kappa$ to be a field for
non-homogeneous materials. $\Delta(x,y)$ is 1 if $|x-y|=1$, 0
otherwise.
In the example we consider a two dimensional spin system but
the extension to arbitrary dimensions is trivial: just change {\tt
mybox}. Moreover we implement a simple {\tt
montecarlo\_multihit} algorithm~\cite{metropolis} 
to minimizes the action in presence of thermal fluctuations.

Here is the parallel program based on {\tt MDP 1.2}.

\begin{verbatim}
// Program: application3.C
#define PARALLEL
#include "MDP_Lib2.h"
#include "MDP_MPI.h"
 
// /////////////////////////////////////////////////////
class scalar_field: public generic_field<float> {
public:
  int ndim,nc;
  scalar_field(generic_lattice &a) {
    allocate_field(a, 1);
    ndim=a.ndim;
  };
  float &operator() (site x) {
    return *(address(x));
  };
};

// /////////////////////////////////////////////////////
class ising_field: public generic_field<int> {
public:
  int ndim,nc;
  float beta, kappa;
  ising_field(generic_lattice &a) {
    allocate_field(a, 1);
    ndim=a.ndim;
  };
  int &operator() (site x) {
    return *(address(x));
  };
  friend void set_cold(ising_field &S) {
    site x(S.lattice());
    forallsites(x) S(x)=1;
  };

  friend void montecarlo_multihit(ising_field &S, scalar_field &H, 
				  int n_iter=10, int n_hits=3) {
    int iter, parity, hit, new_spin, mu;
    float delta_action;
    site x(S.lattice());
    for(iter=0; iter<n_iter; iter++) {
      for(parity=0; parity<=1; parity++)  
        forallsitesofparity(x,parity) {
          for(hit=0; hit<n_hits; hit++) {
            delta_action=S.kappa*H(x);
            for(mu=0; mu<S.ndim; mu++)
              delta_action-=S(x+mu)+S(x-mu);
            new_spin=(S.lattice().random(x).plain()>0.5)?1:-1;
            delta_action*=S.beta*(new_spin-S(x));
            if(exp(-delta_action)>S.lattice().random(x).plain()) 
             S(x)=new_spin;
          };
      };
      S.update(parity);
    };
  };
  friend float average_spin(ising_field &S) {
    float res=0;
    site x(S.lattice());
    forallsites(x) res+=S(x);
    mpi.add(res);
    return res/(S.lattice().nvol_gl);
  };
};

int main(int argc, char **argv) {
  mpi.open_wormholes(argc, argv);
  
  int conf,Ntherm=100,Nconfig=100;
  int mybox[]={64,64};
  generic_lattice mylattice(2, mybox);
  ising_field     S(mylattice); /* ths spin field +1 or -1     */
  scalar_field    H(mylattice); /* the external magnetic field */
  site            x(mylattice);
  JackBoot        jb(Nconfig,1); 
  jb.plain(0);  

  S.beta=0.5;         /* inverse square temperature             */
  S.kappa=0.1;        /* coupling with the total extarnal field */

  /* setting initial conditions */
  forallsites(x) {
    S(x)=1; /* initial spins set to +1 */
    H(x)=0; /* external magnetic field set to zero */
  };
  S.update();
  H.update();

  if(ME==0) 
    printf("Beta\tMagnetization\terror\n"); 
  
  /* looping on values of beta */
  for(S.beta=0.20; S.beta<0.60; S.beta+=0.01) {
    /* termalizing configuration */
    for(conf=0; conf<Ntherm; conf++)
      montecarlo_multihit(S,H);                  
    
    /* generating Nconfig configurations for each value of beta */
    for(conf=0; conf<Nconfig; conf++) {
      montecarlo_multihit(S,H);                  
      jb(conf,0)=average_spin(S);
    };
    if(ME==0) 
      printf("%f\t%f\t%f\n", 
	     S.beta, abs(jb.mean()), jb.j_err());
  };
  mpi.close_wormholes();
  return 0;
};
\end{verbatim} 

\begin{figure}
\epsfxsize=10cm
\epsfysize=10cm
\hfil \epsfbox{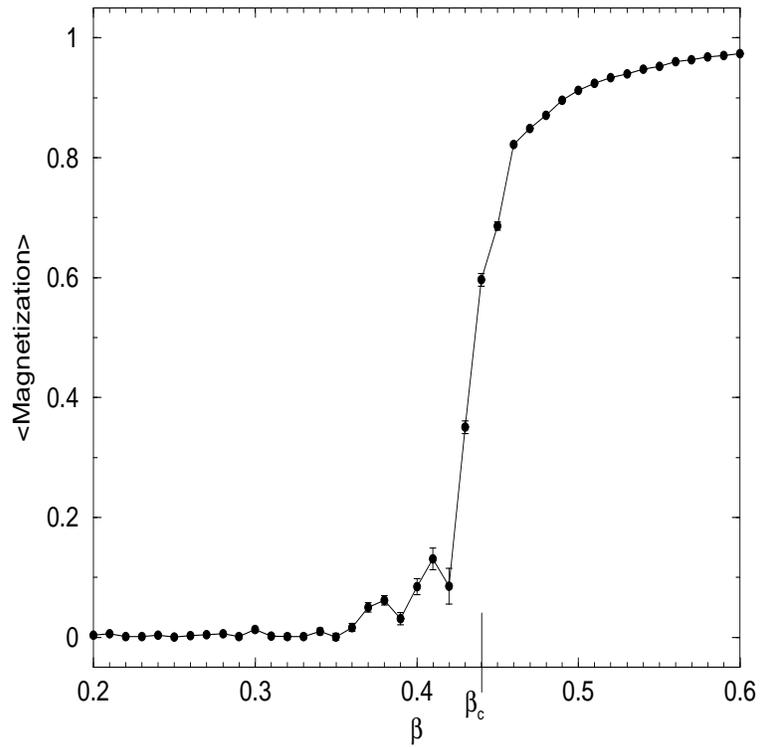} \hfil
\caption{Total average magnetization in a two dimensional spin system as
function of the inverse temperature. The jump corresponds to a phase
transition.} 
\label{isingjump}
\end{figure}

The output of this program (for $h(x)=0$) has been plotted and reported in
fig.~\ref{isingjump}. It clearly shows that when the temperature is
high ($\beta=1/T$ is low) the average magnetization fluctuates around zero
while, when the temperature is low
($\beta$ if high), all the spins tend to become parallel giving origin to a
total magnetization different from zero.
The critical temperature where the phase transition
(the jump) occurs is, for the 2D model 
\begin{equation}
T_c = \frac1{\beta_c} = \frac2{log(1 + \sqrt2)} \simeq 2.269
\end{equation}

Note that the only reference to the number of dimensions is in the 
function {\tt main()} in the lines 
\begin{verbatim}
  int mybox[]={64, 64};
  generic_lattice mylattice(2, mybox);
\end{verbatim}
These are the only lines to change to generalize the program to more than 2D.
For bigger volumes one may also want to change the total number of 
computed configurations {\tt Nconfig} and the steps to thermalization
{\tt Ntherm}. 

\subsection{Solid under stress}

\index{solid under stress}

Many other parallel applications could be developed relatively easy
using {\tt MDP 1.2}.  For example one may be interested in studying
the deformation of a solid under stress. The solid can be approximated
with a finite lattice of given arbitrary shape and each site would be
connected to its next-neighbor and next-next-neighbor sites with
springs. One needs at least 4 fields on this lattice: M,K,G,Q
and P, where M is the mass density of the solid; K the spring
constant; G a parameter determining the internal friction of
the solid; Q the physical (spatial) coordinates of each site, P the
momentum associated to each site. The problem of determining the
deformations of this solid can be solved by iterating a discretized
form of the Hamilton equations
\begin{eqnarray}
P_{t+\Delta t}(x) - P_t(x) &=& \frac{\partial H(P,Q,K,G)}{\partial Q(x)} \Delta t = F(x,Q,P,K,G) \Delta t \\
Q_{t+\Delta t}(x) - Q_t(x) &=& \frac{\partial H(P,Q,K,G)}{\partial P(x)} \Delta t = P(x)/M(x) \Delta t 
\end{eqnarray}
\noindent These equations translates into
\begin{verbatim}
float delta_t=0.000001; /* a small nonzero positive number */ 
forallsites(x)
  for(mu=0; mu<3; mu+) { 
    F[mu]=/* some function of K,G,Q,P */ 
    P(x,mu)=P(x,mu)+F(mu)*delta_t;
    Q(x,mu)=Q(x,mu)+P(x,mu)/m*delta_t; 
  }; 
P.update(); 
Q.update();
\end{verbatim} 

The expression of the function $F$ can be quite complicated depending on
how one models the solid.

\begin{table}
\begin{center}
\begin{tabular}{|l|l|} \hline
{\bf QCD} & {\bf FermiQCD} \\ \hline
\multicolumn{2}{|c|}
{Algebra of Euclidean gamma matrices} \\ \hline
$A=\gamma^{\,\mu} \gamma^{\,5} e^{3 i \gamma^{\,2}}$ &
{\footnotesize\tt Matrix A;} \\
&{\footnotesize\tt A=Gamma[mu]*Gamma5*exp(3*I*Gamma[2]);} \\ \hline
\multicolumn{2}{|c|}
{Multiplication of a fermionic field for a spin structure} \\ \hline
$\forall x: \ \chi(x)=(\gamma^{\,3}+m) \psi(x + \hat \mu)$ &
{\footnotesize\tt /* assuming the following definitions} \\
&{\footnotesize\tt generic\_lattice space\_time(...);} \\
&{\footnotesize\tt fermi\_field chi(space\_time,Nc);} \\
&{\footnotesize\tt fermi\_field psi(space\_time,Nc);} \\
&{\footnotesize\tt site x(space\_time);} \\
&{\footnotesize\tt */} \\
&{\footnotesize\tt forallsites(x)} \\
&{\footnotesize\tt \ \ \ chi(x)=(Gamma[3]+m)*psi(x+mu);} \\ \hline
\multicolumn{2}{|c|}
{Translation of a fermionic field} \\ \hline
$\forall x,a: \ \chi_a(x)=U(x,\mu)\psi_a(x+\hat \mu)$ &
{\footnotesize\tt forallsites(x)} \\
& {\footnotesize\tt \ \ \ for(a=0; a<psi.Nspin; a++)} \\
& {\footnotesize\tt \ \ \ \ \ \ chi(x,a)=U(x,mu)*psi(x+mu,a);} \\ \hline
\end{tabular}
\end{center}
\caption{Example of {\tt FermiQCD} statements.\label{tabFermiQCD}}
\end{table}

\subsection{Lattice QCD} 

\index{Lattice QCD}
\index{\tt FermiQCD}
\index{MILC collaboration}
\index{UKQCD collaboration}
\index{CANOPY}

\label{fermi}

{\tt MDP 1.2} has been used to develop a parallel package for large
scale Lattice QCD simulations called {\tt FermiQCD}~\cite{fermiqcd, 
fermiqcd2}. Here we only list some of its main features.

The typical problem in QCD (Quantum Chromo Dynamics) is that of 
determining the correlation 
functions of the theory as a function of the parameters. 
From the knowledge of these correlation
functions one can extract particle masses and matrix elements to 
compare with experimental results from particle accelerators.

On the lattice, each correlation function is computed numerically 
as the average
of the corresponding operator applied to elements of a Markov chain 
of gauge field configurations.
Both the processes of building the Markov chain and of measuring 
operators involve quasi-local algorithms.
I Some of the main features implemented in {\tt FermiQCD} are:

\begin{itemize}
\item[$\Box \hskip -2.5mm \times$] works on a single process PC 
\item[$\Box \hskip -2.5mm \times$] works in parallel with MPI  
\item[$\Box \hskip -2.5mm \times$] arbitrary lattice partitioning  
\item[$\Box \hskip -2.5mm \times$] parallel I/O (partitioning independent)  
\item[$\Box \hskip -2.5mm \times$] arbitrary space-time dimension  
\item[$\Box \hskip -2.5mm \times$] arbitrary gauge group $SU(n)$  
\item[$\Box \hskip -2.5mm \times$] anisotropic Wilson gauge action  
\item[$\Box \hskip -2.5mm \times$] anisotropic Wilson fermionic action  
\item[$\Box \hskip -2.5mm \times$] anisotropic Clover improved action  
\item[$\Box \hskip -2.5mm \times$] D234 improved action  
\item[$\Box \hskip -2.5mm \times$] Kogut-Susskind improved action  
\item[$\Box \hskip -2.5mm \times$] ordinary and stochastic propagators  
\item[$\Box \hskip -2.5mm \times$] minimal residue inversion  
\item[$\Box \hskip -2.5mm \times$] stabilized biconjugate gradient inversion  
\item[$\Box \hskip -2.5mm \times$] twisted boundary conditions for large $\beta$ numerical perturbation theory (all fields) 
\item[$\Box \hskip -2.5mm \times$] reads existing CANOPY data  
\item[$\Box \hskip -2.5mm \times$] reads existing UKQCD data  
\item[$\Box \hskip -2.5mm \times$] reads existing MILC data  
\end{itemize}

In table~\ref{tabFermiQCD} we show a few examples of 
FermiQCD Object Oriented 
capabilities (compared with examples in the standard 
textbook notation for Lattice QCD~\cite{fermiqcd2})

\section{Timing and efficiency issues}

\begin{figure}
\epsfxsize=10cm
\hfil \epsfbox{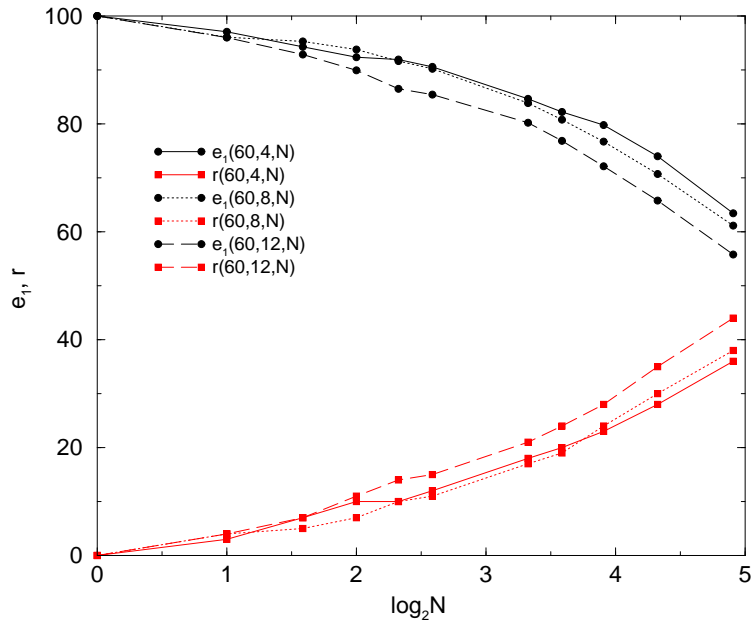} \hfil
\caption{Efficiency $e_1$ as function of $N$ and $L$, for $T=60$ 
(the number 60 has been chosen because it has the biggest number 
of integer factors within the number of available processors).}
\label{efficiency1}
\end{figure}

\begin{figure}
\epsfxsize=10cm
\hfil \epsfbox{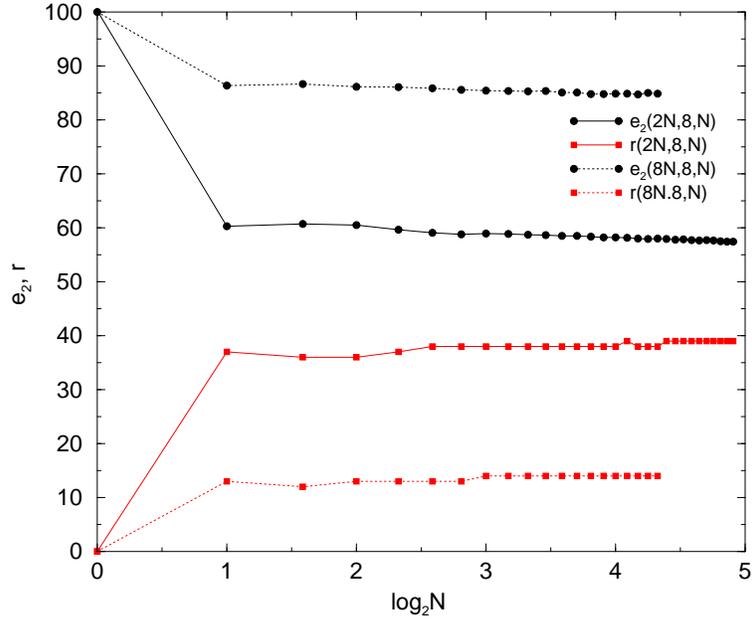} \hfil
\caption{Efficiency $e_2$ as function of $N$, for $T=2N$ and $L=8$.}
\label{efficiency2}
\end{figure}

The issue of efficiency of a parallel program is not a simple one,
in fact there are many factors to take into account. In particular
the efficiency scales with the number of parallel processes, with
the size of the data that need to be communicated between processes,
and with the total volume of the lattice.

The dependence of the efficiency on these variables cannot
be in general factorized because of non-linearities introduced 
by modern 
computer architectures (for example caches effects, 
finite memory, memory speed/cpu speed, latencies, etc).

To give an idea of how a parallel computer program based on {\tt MDP} 
scales with the number of processes, we consider the following program.

\begin{verbatim}
// program: laplace.C
#define PARALLEL
#include "MDP_Lib2.h"
#include "MDP_MPI.h"
 
int main(int argc, char **argv) {
  mpi.open_wormholes(argc, argv);
  int box[4]={8,12,12,12};
  generic_lattice space(4,box);
  Matrix_field U(space,3,3);
  Matrix_field V(space,3,3);
  site x(space);
  forallsites(x) {
    V(x)=space.random(x).SU(3);
    U(x)=0;
  };
  U.update();
  V.update();
  for(int i=0; i<10000; i++) {
    forallsites(x)
      U(x)=0.125*(cos(U(x)+V(x))+
                  U(x+0)+U(x-0)+
                  U(x+1)+U(x-1)+
                  U(x+2)+U(x-2)+
                  U(x+3)+U(x-3));
    /* uncomment to print convergence
    if(ME==0) {
      x.set(0,0,0,0);
      printf("%f\n", real(U(x,0,0)));
    };
    */
    U.update();
  };
  V.save("V_field.dat");
  U.save("U_field.dat");
  mpi.close_wormholes();
  return 0;
}; 
\end{verbatim}

It solves a 4-dimensional Laplace equation for a field of $3 \times 3$ 
matrices, and performs some parallel IO as well. The reader can work out 
the details of the program as an exercise.

We ran the same program for different lattice volumes $V=T\times L^3$
and on different sets of processors ($N$=number of processors).
Each processor (node) is a 700MHz PentiumIII PC. The nodes are connected 
using Myrinet cards/switches.

For each run we measured the total time $t(T,L,N)$ and we then computed 
the efficiency according with two possible definitions%
\footnote{In the context of Lattice QCD simulation, up to an 
overall normalization, the first definition
is the one used at Fermilab to benchmark parallel software, 
the second is the definition used by the MILC collaboration.}:
\begin{eqnarray}
e_1(T,L) &=& 100 \frac{t(T,L,1)}{N\,t(T,L,N)} \\
e_2(N,L,x) &=& 100 \frac{t(x,L,1)}{t(N x,L,N)}
\end{eqnarray}
The first definition is applicable to computations in which the 
problem size is fixed and one varies the numbers of processors.
The second definition is applicable to computations for which the
size of the problem increases with the number of processors available.
In both cases the efficiency is normalized to $100\%$.
In an ideal world (without latencies and with instantaneous communications)
both these efficiencies should be constant and equal to 100\%.

In figures~\ref{efficiency1}-\ref{efficiency2} we plot $e_1$ 
and $e_2$ for our runs and the maximum time spent in communications by the nodes (as computed by {\tt MDP}) that we call $r$. The normalizations are given by
\begin{eqnarray}
t(60,4,1)  &=& 788\text{sec.} \nonumber \\
t(60,8,1)  &=& 6417\text{sec.} \nonumber \\
t(60,12,1) &=& 6417\text{sec.} \nonumber \\
t(2,8,1) &=& 201\text{sec.} \nonumber \\
t(8,8,1) &=& 834\text{sec.} \nonumber
\end{eqnarray}
These numbers can be further reduces by writing a user 
defined function that optimizes the critical computation done 
in the line
\begin{verbatim}
      U(x)=0.125*(cos(U(x)+V(x))+ ...
\end{verbatim}

From the plots we observe that with good approximation
\begin{equation}
e_i \simeq 100 - r
\end{equation}
for both the definitions of efficiency.

Eventually, for $N$ bigger than some $N_0$, the total time
becomes totally dominated by communication 
(the time spent in communications goes to 100\%) 
and $e_1$ starts to scale as $1/N$. 

Note that the answer to the question ``how many processors should I use?'' 
is not addressed in this paper because, in financial terms, 
this depends on the subjective utility function of each individual user.
The only general conclusion is that $N$ should be less than $N_0$.

A different algorithm may scale in a different way even if 
the qualitative behavior of the efficiency should be the same. Some 
algorithms may require more communications than others and may 
result in a bigger loss of efficiency when running in parallel.

\section*{Acknowledgements}

I thank Chris Sachrajda,
Jonathan Flynn and Luigi Del Debbio of the University of Southampton
(UK) where the {\tt MDP} project started.

The development of the code {\tt MDP 1.2} in its final (but not yet
definitive) form and its main application, {\tt FermiQCD},
have greatly benefit from discussions and suggestions
with members of the Fermilab Theory Group. In particular I want to
thank Paul Mackenzie, Jim Simone, Jimmy Juge, Andreas Kronfeld and
Estia Eichten for comments and suggestions.  
Moreover I freely borrowed many ideas from existing Lattice QCD codes 
(including the UKQCD code~\cite{ukqcd}, the MILC code~\cite{milc}, 
QCDF90~\cite{qcdf90} and CANOPY~\cite{canopy}). 
I here thank their authors for making their software available to me.
\vskip 5mm
The realization of this paper and of the software {\tt FermiQCD}
was performed at Fermilab, a U.S. Department of Energy Lab 
(operated by the University Research Association, Inc.), 
under contract DE-AC02-76CHO3000.

\newpage

\begin{center}
{\LARGE LICENSE for {\tt MDP 1.2}}

(including examples and applications such as {\tt FermiQCD}).
\end{center}
\begin{itemize}
{\footnotesize

\item {\tt MDP 1.2} has been created by Massimo Di Pierro. {\tt MDP
1.2} is a property of the author. (The application {\tt FermiQCD}, 
also covered by this License is a joined property of the author and
of Fermilab).

\item {\tt MDP 1.2} is free of charge for educational and research 
institutions worldwide. 

\item You may copy and distribute exact replicas of {\tt MDP 1.2} 
as you receive it, 
in any medium, provided that you conspicuously and appropriately 
publish on each copy an appropriate copyright notice including 
the author's name and the disclaimer of warranty; 
keep intact all the notices that refer to this License and to the 
absence of any warranty; and give any other recipient of {\tt MDP 1.2} 
a copy of this License along with the software.  

\item You may modify your copy or copies of {\tt MDP 1.2} or any
portion of it and distribute such modifications or work under the
terms of Section 1 and 2 above, provided that the modified content 
carries prominent notices stating that it has been changed, the exact 
nature and content of the changes, and the date of any change. 
 
\item BECAUSE {\tt MDP 1.2} IS LICENSED FREE OF CHARGE, THERE IS
NO WARRANTY FOR IT. EXCEPT WHEN OTHERWISE STATED IN WRITING.
THE AUTHOR PROVIDES {\tt MDP 1.2} "AS IS" WITHOUT WARRANTY OF 
ANY KIND, EITHER EXPRESSED OR IMPLIED.
THE ENTIRE RISK OF USE OF THIS SOFTWARE IS WITH THE RECIPIENT. 
SHOULD {\tt MDP 1.2} OR ONE OF ITS APPLICATIONS PROVE FAULTY, INACCURATE, 
OR OTHERWISE UNACCEPTABLE YOU ASSUMES THE COST OF ALL NECESSARY 
REPAIR OR CORRECTION.  

\item To request {\tt MDP 1.2}
(including the examples and applications 
described in this paper) please sign this license and 
fax or mail this page to the author:

\begin{verbatim}
	Massimo Di Pierro             
	Email:   mdp@fnal.gov
	Fax:     001-630-840-5435
	Address: MS 106, Fermilab, Batavia, IL 60510-0500 (USA)
\end{verbatim}}
\end{itemize}

\hskip 2cm \vbox{\begin{verbatim}
Name: .................................
Date: .................................
Email: ................................
Signiture: ............................
\end{verbatim}}

\appendix

\section{Classes declared in {\tt MDP\_MPI.h}}

\begin{verbatim}
#define NOWHERE 268435456
#define EVEN 0
#define ODD 1
#define EVENODD 2


#define ME mpi.return_id()
#define _NprocMax_ 256
#ifndef Nproc
int Nproc=1; 
#endif

#define forallsites(x)
#define forallsitesofparity(x,parity)
#define forallsitesandcopies(x)
#define forallsitesandcopiesofparity(x,parity)


class mpi_wormhole_class {
open_wormholes(int&, char***);
close_wormholes();
template <class T> void put(T& object, int destination);
template <class T> void put(T& object, int destination, 
                            mpi_request&);
template <class T> void get(T& object, int source);
template <class T> void put(T* object, long size, int destination);
template <class T> void put(T* object, long size, int destination, 
                            mpi_request&);
template <class T> void get(T* object, long size, int source);
void wait(mpi_request&);
void wait(mpi_request*, int size);
void add(float& source_obj, float& dest_obj);
void add(float* source_obj, float* dest_obj, long size);
void add(double& source_obj, double& dest_obj);
void add(couble* source_obj, double* dest_obj, long size);
void add(int&);
void add(long&);
void add(float&);
void add(double&);
void add(Complex&);
void add(Matrix&);
void add(int&,long);
void add(long&,long);
void add(float&,long);
void add(double&,long);
void add(Complex&,long);
void add(Matrix&,long);
template <class T> void broadcast(T& object, int source);
template <class T> void broadcast(T* object, long size, int source);
void barrier();
double time();
void reset_time();
wait abort();
} mpi; 

class generic_lattice {
int  ndim;
long nvol;
generic_lattice(int ndim, 
                int *nx,
                int (*where)(int*,int,int*)=
                    default_partitioning<0>,
                void (*topology)(int,int*,int*,int*,int,int*)=
                    torus_topology,
                long seed=0, 
                int  next_next=1);
generic_lattice(int ndim, 
                int ndir,
                int *nx,
                int (*where)(int*,int,int*)=
                    default_partitioning,
                void (*topology)(int,int*,int*,int*,int,int*)=
                    torus_topology,
                long seed=0, 
                int  next_next=1);
void allocate_lattice(int ndim, 
                int *nx,
                int (*where)(int*,int,int*)=
                    default_partitioning,
                void (*topology)(int,int*,int*,int*,int,int*)=
                    torus_topology,
                long seed=0, 
                int  next_next=1);
void allocate_lattice(int ndim,
                int ndir, 
                int *nx,
                int (*where)(int*,int,int*)=
                    default_partitioning,
                void (*topology)(int,int*,int*,int*,int,int*)=
                    torus_topology,
                long seed=0, 
                int  next_next=1);
long global_coordinate(int *);
void global_coordinate(long, int *);
int  compute_parity(int*);
void deallocate_memory();
void intitialize_random(long seed);
Random_generator& random(site x);
long local_volume();
long global_volume();
int n_dimensions();
int n_directions();
long size();
long size(int);
long local_volume();
long global_volume();
long move_up(long, int);
long move_down(long,int);
long start_index(int, int);
long stop_index(int,int);
};

int on_which_process(generic_lattice &lattice, 
                     int x0, int x1, int x2, ...);

class vector {
   int x[10];
};

vector bynary2versor(long);
int    versor2binary(int,int,int,int,int...);
long   vector2binary(vector);

class site {
site(generic_lattice& mylattice);
void start(int parity);
void next();
int is_in();
int is_here();
int is_in_boundary();
int is_equal(int x0, int x1, int x2, ...);
int parity();
long local_index();
long global_index();
void set(int x0, int x1, int x2, ...);
void set_local(long index);
void set_global(long index);
site operator+(int mu);
site operator-(int mu);
int operator()(int mu);
int operator=(site x);
int operator=(int *x);
int operator==(site x);
int operator==(int *x);
int operator!=(site x);
int operator!=(int *x);
int operator=(vector);
int operator+(vector);
int operator-(vector);
};

long site2binary(site);
 
template<class T>
class generic_field {
generic_field();
generic_field(generic_lattice& mylattice, 
	      int field_components);
void allocate_field(generic_lattice& mylattice,
              int field_components);
void deallocate_memory();
T* address(site x);
T* address(site x, int field_component);
void operator=(generic_field& psi);
generic_lattice& lattice();
void operator=(const generic_field&);
void operator=(const T);
void operator+=(const generic_field&);
void operator==(const generic_field&);
void operator+=(const generic_field&);
template<class T2>
void operator*=(const T2);
template<class T2>
void operator/=(const T2);
void update(int parity, int block_n, int block_size);
void load(char filename[], int processIO, long buffersize);
void save(char filename[], int processIO, long buffersize);
long physical_size();
long size_per_site();
long physical_local_start(int);
long physical_local_stop(int);
T* physical_address()
T& physical_address(long)
void assign_to_null();
};

class Matrix_field: public generic_field<Complex> {
Matrix_field(generic_lattice& mylattice, 
             int rows, 
             int columns);
Matrix operator() (site x);
Complex& operator() (site x, int row, int columns);
};

class NMatrix_field: public generic_field<Complex> {
Matrix_field(generic_lattice& mylattice, 
             int n_matrices,
             int rows, 
             int columns);
Matrix operator() (site x, int n_matrix);
Complex& operator() (site x, int n_matrix 
                     int row, int columns);
};

class Vector_field:  public generic_field<Complex> {
Vector_field(generic_lattice& mylattice,
             int elements);
Matrix operator() (site x);
Complex& operator() (site x, int element);
};

class NVector_field:  public generic_field<Complex> {
NVector_field(generic_lattice& mylattice,
              int n_vectors,
              int elements);
Matrix operator() (site x, int n_vector);
Complex& operator() (site x, int n_vector, int element);
};
\end{verbatim} 

\end{document}